\pdfoutput=1
\documentclass{article}
\PassOptionsToPackage{numbers, sort&compress}{natbib}
\usepackage[preprint]{neurips_2025}
\usepackage[T1]{fontenc}
\usepackage{xcolor}
\usepackage[colorlinks=true, linkcolor=blue!60!black, citecolor=blue!60!black, urlcolor=blue!60!black]{hyperref}
\hypersetup{pdftitle={Variants of Beam Search for DiskANN}, pdfauthor={Viraj Agrawal, Kirankumar Shiragur}}
\usepackage{url}
\usepackage{booktabs}
\usepackage{amssymb}
\usepackage{graphicx}
\usepackage{microtype}

\usepackage{amsmath}
\usepackage{amsthm}
\usepackage{algorithm}
\usepackage{algorithmic}
\usepackage{enumitem}
\usepackage{mdframed}
\usepackage{subcaption}
\usepackage{multirow}
\usepackage{placeins}

\newsavebox{\fitbox}
\newcommand{\fitwidth}[1]{\sbox{\fitbox}{#1}\ifdim\wd\fitbox>\linewidth\resizebox{\linewidth}{!}{\usebox{\fitbox}}\else\usebox{\fitbox}\fi}

\newcommand{\nout}[1]{N_{\mathrm{out}}\!\left(#1\right)}
\newcommand{\Rbb}{\mathbb{R}}

\theoremstyle{plain}
\newtheorem{theorem}{Theorem}[section]
\theoremstyle{definition}
\newtheorem{definition}[theorem]{Definition}

\theoremstyle{plain}

\newmdenv[
  linecolor=black!20,
  backgroundcolor=black!3,
  linewidth=0.5pt,
  innerleftmargin=8pt,
  innerrightmargin=8pt,
  innertopmargin=6pt,
  innerbottommargin=6pt,
  skipabove=6pt,
  skipbelow=6pt,
]{softbox}

\title{Variants of Beam Search for DiskANN}

\author{%
  Viraj Agrawal\\
  Microsoft Research\\
  India\\
  \texttt{t-viragrawal@microsoft.com}
  \And
  Kirankumar Shiragur\\
  Microsoft Research\\
  India\\
  \texttt{kshiragur@microsoft.com}
}

\begin{document}

\maketitle

\begin{abstract}
    Graph-based data structures such as DiskANN, HNSW, and NSG constitute the state of the art for large-scale approximate nearest-neighbor (ANN) search. Indyk and Xu~\cite{indykxu} introduced the notion of $\alpha$-reachable graphs, which include graphs built by DiskANN's slow-preprocessing construction, and proved that greedy search achieves an approximation factor of $\tfrac{\alpha+1}{\alpha-1}$. Independently, Al-Jazzazi et al.~\cite{adaptivebeam} proposed adaptive beam search (ABS) with a relaxation parameter $\gamma > 0$, showing a $\tfrac{2}{\gamma}$ approximation on navigable graphs ($\alpha \approx 1$), where greedy search admits no guarantees. In this work, we unify and generalize these results by proving that ABS with parameter $\gamma$ on $\alpha$-reachable graphs achieves an approximation of $\tfrac{\alpha+1}{\alpha(\gamma+1)-1}$ for any metric. Moreover, on sorted $\alpha$-reachable graphs (introduced by Gollapudi et al.~\cite{sortprune}) under the $\ell_2$ metric, we obtain the tighter bound $\max\bigl(\tfrac{\alpha}{\alpha(\gamma+1)-1}, \tfrac{\alpha^{2}+1}{\alpha^{2}(\gamma+1)-1}\bigr)$. Beyond these theoretical results, we design a beam predictor using features inspired by fine-grained analysis of ABS, trained via a constrained objective that minimizes mean distance computations subject to a target recall. On three 100-million-point benchmarks, using graphs built by DiskANN's practical construction, beam predictor lowers mean query latency on SSD-resident indices at matched recall by up to $29.6\%$ relative to classical beam search (CBS). Relative to ABS, it is faster in $11$ of $15$ settings, by up to $6.5\%$, and at most $1.4\%$ slower in the other $4$; at $95\%$ recall, its advantage over ABS grows from $0.8$--$4.2\%$ for $100$ neighbors to $4.6$--$5.6\%$ for $500$ neighbors. In memory, it needs $1.3$--$23.7\%$ fewer distance computations than CBS and $0.0$--$3.9\%$ fewer than ABS.
\end{abstract}

\section{Introduction}\label{sec:intro}

Nearest neighbor search is a fundamental problem in data structures, with applications in search, recommendation systems and, more recently, retrieval-augmented generation (RAG). Given a dataset of $n$ vectors and a query vector $q$, the goal is to retrieve the $k$ data vectors nearest to $q$. Since exact search typically takes linear time in $n$ in high dimensions, much research has focused on \emph{approximate} nearest neighbor (ANN) search: data structures and algorithms that efficiently return high-quality approximate answers. Early work on ANN search targeted modest datasets---millions of points in tens of dimensions---and produced a rich line of space-partitioning techniques, including locality-sensitive hashing~\cite{indyk1998approximate, andoni2008near}, KD-trees~\cite{arya1998optimal} and cover trees~\cite{Beygelzimer06}. These methods scale poorly with the dimension: tree-based partitions must examine a number of cells that grows exponentially with it, and hashing needs many hash tables to reach high recall. To avoid this curse of dimensionality, more recent work has turned to graph-based indices such as DiskANN~\cite{diskann}, HNSW~\cite{hnsw}, NSG~\cite{nsg} and NGT~\cite{NGT1,NGT2,NGT3}, which have become the dominant approach in practice due to their strong empirical performance.

Despite this empirical success, the theoretical understanding of graph-based methods remained limited until the recent work of Indyk and Xu~\cite{indykxu}, who introduced a family of graphs called $\alpha$-reachable graphs ($\alpha > 1$). This family includes the graphs built by DiskANN's slow preprocessing, and the authors showed that a simple greedy search on such graphs converges to a local optimum whose distance to the query $q$ is at most $\tfrac{\alpha+1}{\alpha-1}$ times the true nearest-neighbor distance. Subsequently, Gollapudi et al.~\cite{sortprune} extended this analysis to the widely used classical beam search (CBS), showing the same $\tfrac{\alpha+1}{\alpha-1}$ approximation guarantee for all elements maintained in the beam. In addition, they showed that DiskANN’s RobustPrune procedure produces a special class of \emph{sorted} $\alpha$-reachable graphs, and proved that under the $\ell_2$ metric these graphs admit the sharper approximation bound of $\tfrac{\alpha}{\alpha-1}$. Complementing these works, Al-Jazzazi et al.~\cite{adaptivebeam} proposed a modified search procedure called \emph{adaptive beam search} (ABS), parameterized by $\gamma > 0$, which delays termination to enable additional exploration. They proved a $\tfrac{2}{\gamma}$-approximation guarantee on \emph{navigable} graphs—corresponding in our terminology to the regime $\alpha \approx 1$—where classical beam search does not admit any provable guarantees.

Given these two extremes—classical beam search on $\alpha$-reachable graphs $(\alpha > 1, \gamma = 0)$ and adaptive beam search on navigable graphs $(\alpha \approx 1, \gamma > 0)$—it is natural to ask the following questions.

\begin{softbox}
\textbf{Q1.} What approximation guarantee can adaptive beam search, parameterized by $\gamma \geq 0$, achieve on $\alpha$-reachable graphs with $\alpha > 1$?

\end{softbox}

\begin{softbox}
\textbf{Q2.} Can the improved $\ell_2$-metric guarantee of Gollapudi et al.~\cite{sortprune} be combined with adaptive beam search?
\end{softbox}

\paragraph{Our results.}
We answer both questions and, building on the intuition behind ABS, design beam predictor, which chooses the list size for each query independently for the search; in our experiments it is faster than CBS in $14$ of the $15$ SSD settings and within the run-to-run variation of it in the other, and faster than ABS in most settings. Our first result addresses the first question by showing that adaptive beam search on $\alpha$-reachable graphs achieves an improved approximation guarantee, summarized in the following theorem.

\begin{theorem}\label{thm:main1}
Let $G=(V,E)$ be an $\alpha$-reachable graph with $\alpha>1$, let $q$ be a query whose distances to the points of $V$ are pairwise distinct, let $B^{\star}=\{b_1,\dots,b_k\}$ be the result of adaptive beam search (sorted by increasing distance from $q$) with parameter $0 \le \gamma \le 2/\alpha$, and let $A = \{a_1,\dots,a_k\}$ denote the true top-$k$ neighbors of $q$ sorted by increasing distance. Then for any metric $D$,
\[
D(b_i,q) \;\le\; C \cdot D(a_i,q), \quad i=1,\dots,k,
\qquad
C = \frac{\alpha+1}{\alpha(\gamma+1)-1}.
\]
\end{theorem}

Building on this result, our second contribution addresses the second question by establishing a sharper approximation guarantee for adaptive beam search on sorted $\alpha$-reachable graphs under the $\ell_2$ metric, as summarized below.

\begin{theorem}\label{thm:main2}
Under the hypotheses of \autoref{thm:main1}, if in addition $G$ is \emph{sorted} $\alpha$-reachable and $D$ is the $\ell_2$ metric, then for any $0 \le \gamma \le \max(1/\alpha,2/\alpha^{2})$,
\[
D(b_i,q) \;\le\; C \cdot D(a_i,q), \quad i=1,\dots,k,
\]
\[
C = \max\!\left(\frac{\alpha}{\alpha(\gamma+1)-1},\frac{\alpha^{2}+1}{\alpha^{2}(\gamma+1)-1}\right).
\]
\end{theorem}

The general-metric bound is tight, and it recovers the bound of Indyk and Xu~\cite{indykxu} at $\gamma = 0$ and that of Al-Jazzazi et al.~\cite{adaptivebeam} as $\alpha \to 1$. The sorted-$\ell_2$ bound is tight where the first term of the $\max$ dominates, and it recovers the bound of Gollapudi et al.~\cite{sortprune} at $\gamma = 0$. \autoref{sec:theory} proves both theorems, with the semidefinite program used for \autoref{thm:main2} solved in Appendix~\ref{app:thm:mainproof}, and gives the tight examples (\autoref{sec:tight}).

Both theorems concern (sorted) $\alpha$-reachable graphs, such as those built by DiskANN's slow preprocessing, whose naive construction takes $O(n^3)$ time. The graphs that DiskANN builds in practice, including those in our experiments, are not known to satisfy these definitions, so the guarantees do not cover them. However, on these graphs used in practice, recall moves in the direction that the bounds suggest: on three 100-million-point datasets, the recall of ABS rises with $\gamma$, and it is higher on a graph built with DiskANN's sorted pruning rule than on one built with the candidates in random order (\autoref{sec:exp-theory}).

Why does ABS help? As Al-Jazzazi et al.~\cite{adaptivebeam} observe, it adapts to the difficulty of each query. Its stopping rule compares the next candidate with the query's current top-$k$, so it stops early on an easy query, whose top-$k$ settles after a few steps, and keeps searching a hard query, whose top-$k$ is still improving. In effect, ABS moves distance computations away from queries that are already accurate, where more search buys little recall, toward queries that are not, where it buys the most. This is how it matches the mean recall of CBS with fewer distance computations, whereas CBS gives every query the same list size.

\textbf{Beam predictor.} We make this reallocation explicit and optimize it. Beam predictor chooses a list size for each query: it first runs a short pilot beam search, extracts features from the nodes it visits and their distances to the query, predicts from these features a list size that reflects how close the query is to convergence, and continues the search with that list size instead of starting over. It is trained to reproduce the list sizes that minimize the total number of distance computations of a batch of queries subject to a target mean recall, an allocation that we compute with a Lagrangian relaxation. Per-query search budgets have been studied before (\autoref{sec:related}); our beam predictor differs from those methods in learning from this cost-optimal allocation.

On three 100-million-point benchmarks from the NeurIPS'21 Billion-Scale ANN Challenge~\cite{bigann-benchmarks}, with search done on graphs built with DiskANN's sorted pruning rule, beam predictor has a lower mean latency than CBS in $14$ of the $15$ SSD settings that we test, by up to $29.6\%$, and in the remaining one it is $0.6\%$ slower, within the run-to-run variation. Compared with ABS, it is faster in $11$ of the $15$ settings, by up to $6.5\%$, and at most $1.4\%$ slower in the other $4$; in memory, it needs $0.0$--$3.9\%$ fewer distance computations than ABS. On the SSD at $95\%$ recall, its lead over ABS is larger at $K = 500$ than at $K = 100$ on all three datasets: from $K=100$ to $K=500$, the speedup of ABS over CBS shrinks from $10.4\%$ to $1.7\%$ on SIFT100M, while that of beam predictor shrinks less, from $14.2\%$ to $7.2\%$, and at $K=500$ beam predictor is the fastest of the three methods on all three datasets. These gains are in mean latency: beam predictor gives long searches to the queries that it predicts to be hard, and short searches to the queries that it predicts to be easy.

\paragraph{Contributions.} In summary, we make the following contributions.
\begin{itemize}[leftmargin=*, itemsep=1pt, topsep=2pt]
\item \textbf{Guarantees for ABS on $\alpha$-reachable graphs.} On $\alpha$-reachable graphs, ABS with parameter $\gamma$ returns a $\tfrac{\alpha+1}{\alpha(\gamma+1)-1}$-approximation for any metric (\autoref{thm:main1}), and on sorted $\alpha$-reachable graphs under $\ell_2$ it achieves the sharper bound of \autoref{thm:main2}. The bounds are tight, the second where its first term dominates, and they recover those of Indyk and Xu, Gollapudi et al.\ and Al-Jazzazi et al.\ as special cases (\autoref{sec:theory}).
\item \textbf{Beam predictor.} A per-query list-size predictor that is trained on the cost-optimal allocation of list sizes for a target mean recall and continues the short pilot search from which it extracts its features instead of restarting the entire search, so that the pilot's work is reused (\autoref{sec:predl-main}).
\item \textbf{Evaluation.} An evaluation on three 100-million-point benchmarks, on an SSD-resident index and in memory, both on graphs built with DiskANN's sorted pruning rule, showing that beam predictor is faster than CBS in $14$ of $15$ SSD settings (and within the run-to-run variation of it in the other) and faster than ABS in $11$ of $15$ SSD settings (and at most $1.4\%$ slower in the other $4$), with a lead over ABS on the SSD at $95\%$ recall that is larger at $K = 500$ than at $K = 100$ (\autoref{sec:experiments}).
\end{itemize}

\section{Related work}\label{sec:related}

\paragraph{Graph-based search and its theory.}
Graph-based indices such as HNSW~\cite{hnsw}, NSG~\cite{nsg} and DiskANN~\cite{diskann} perform strongly in large-scale ANN benchmarks~\cite{bigann-benchmarks,simhadri2024results}, and all three answer queries with beam search under a fixed list size. DiskANN, FreshDiskANN~\cite{freshdiskann} and SPANN~\cite{SPANN} serve billion-point datasets from a single machine with an SSD. Early work on graph-based search assumes random or low-dimensional data: Laarhoven~\cite{laarhoven2017graph} derives time-space trade-offs for greedy walks on near-neighbor graphs of random points on the sphere, and Prokhorenkova and Shekhovtsov~\cite{prokhorenkova2020graph} analyze greedy search, shortcut edges and candidate lists when $d \ll \log n$. Without distributional assumptions, Diwan et al.~\cite{diwan2024navigable} bound how sparse a navigable graph can be, Indyk and Xu~\cite{indykxu} prove an approximation guarantee for greedy search on $\alpha$-reachable graphs, Gollapudi et al.~\cite{sortprune} extend it to beam search and tighten it on sorted $\alpha$-reachable graphs under $\ell_2$, and Al-Jazzazi et al.~\cite{adaptivebeam} prove an approximation guarantee for ABS on any navigable graph. Our bounds combine the last three results. Other graph constructions come with guarantees for search on them, typically under assumptions on the query or the data, among them MRNG~\cite{nsg}, FANNG~\cite{fanng}, $\tau$-MG~\cite{taumg}, $\alpha$-CG~\cite{alphacg} and LMG~\cite{lmg}; these guarantees concern their own constructions and search procedures, whereas ours concern ABS on (sorted) $\alpha$-reachable graphs.

\paragraph{Adapting the search effort to each query.}
Queries differ widely in how much search they need; this difficulty has been related to local intrinsic dimensionality~\cite{Aumuller} and to graph-native measures such as Steiner-hardness~\cite{steinerhardness}. Several methods therefore choose the search effort per query. LAET~\cite{li2020improving} predicts a termination condition for each query from features collected during the search, and Tao~\cite{tao} predicts it before the search from static features of the query. DARTH~\cite{darth} terminates the search once a learned model estimates that the query has reached a declared recall target, PiP~\cite{pip} terminates it once the improvement of the current nearest neighbors falls below a threshold, and Ada-ef~\cite{adaef} sets the HNSW list size of each query from a statistical model of the similarities between the query and the data. ANNiE~\cite{annie} runs a preliminary search towards the query and predicts, from the neighborhood it reaches, the cost of answering the query at a given recall target. Beam predictor likewise predicts a per-query list size from a short pilot search, but it learns from the list sizes that minimize the total cost of a batch of queries subject to a target mean recall. This allocation is a multiple-choice knapsack problem, which we solve by Lagrangian relaxation.

\section{Preliminaries}\label{sec:prelim}

We consider a metric space $(\mathcal{X},D)$ where $\mathcal{X}$ is a set of points and $D: \mathcal{X} \times \mathcal{X} \to \Rbb_{\ge 0}$ is a distance function that satisfies, for all $u,v,p \in \mathcal{X}$: (i) $D(u,v) \ge 0$; (ii) $D(u,v)=0 \iff u = v$; (iii) $D(u,v) = D(v,u)$; and (iv) the triangle inequality $D(u,p) \le D(u,v) + D(v,p)$. We are given a dataset of $n$ points in $\mathcal{X}$, which form the node set $V$ of a directed graph $G=(V,E)$, and we let $\nout{u} = \{v \in V : (u,v) \in E\}$ be the out-neighbors of $u$. The definition of $\alpha$-reachable is as follows.

\begin{algorithm}[tb]
  \caption{General beam search (adapted from~\cite{adaptivebeam})}
  \label{alg:beamsearch}
  \begin{algorithmic}[1]
    \STATE {\bfseries Input:} Graph $G=(V,E)$, start node $s \in V$, distance function $D$, query $q \in \mathcal{X}$, target $k \le |V|$.
    \STATE {\bfseries Output:} Set $B^{\star} = \{b_1,\dots,b_k\}$ of $k$ approximate nearest neighbors of $q$.
    \STATE Initialize two min-priority queues (sorted by distance to $q$): $B$ containing all discovered nodes, $B_u \subseteq B$ containing the not-yet-expanded nodes.
    \STATE Insert $s$ into both $B$ and $B_u$; set $B^{\star} \gets \emptyset$.
    \WHILE{$B_u \neq \emptyset$}
      \STATE $x \gets$ top element of $B_u$.
      \IF{\textbf{termination condition} holds}
        \STATE \textbf{break}.
      \ENDIF
      \STATE Pop $x$ from $B_u$.
      \FOR{$y \in \nout{x}$}
        \IF{$y \notin B$}
          \STATE Insert $y$ into $B$ and $B_u$.
        \ENDIF
      \ENDFOR
    \ENDWHILE
    \STATE Return the top $k$ elements of $B$ as $B^{\star}$.
  \end{algorithmic}
\end{algorithm}

\begin{definition}[$\alpha$-reachable graph~{\cite[Definition~3.1]{indykxu}}]
\label{def:alpha}
A directed graph $G=(V,E)$ is \emph{$\alpha$-reachable} with $\alpha > 1$ if for every pair $u,v \in V$ and $u \neq v$, either $(u,v) \in E$ or there exists $t \in V$ such that $(u,t) \in E$ and $D(v,t) \le D(v,u)/\alpha$.
\end{definition}

Informally, from any node $u$ there is a local step towards any target $v$ whose distance shrinks by a factor of at least $\alpha$.

\textbf{DiskANN graphs.} DiskANN~\cite{diskann} builds its graph with the Vamana algorithm, whose edge selection uses the RobustPrune algorithm described below. Following~\cite{indykxu}, we distinguish two constructions built on this algorithm. \emph{Slow preprocessing} prunes each point's out-edges against all other points, with no degree bound; a naive implementation takes $O(n^3)$ time, and the result is an $\alpha$-reachable graph. \emph{Fast preprocessing} is Vamana itself: it prunes only the candidates found by greedy search on the partially built graph and caps the out-degree at $R$. It carries no such guarantee. Our guarantees (\autoref{sec:theory}) hold for any (sorted) $\alpha$-reachable graph, and in particular for slow-preprocessing graphs. Our experiments use fast-preprocessing graphs, the construction used in practice (parameters in \autoref{sec:experiments}).

Both variants rely on the \emph{RobustPrune} subroutine~(Algorithm~2 of~\cite{diskann}). Given a node $p$, a candidate set $\mathcal{V}$, a parameter $\alpha$, and a degree bound $R$, RobustPrune first \emph{sorts} $\mathcal{V}$ in increasing distance from $p$, then iterates through the sorted list and adds each $p'$ to $\nout{p}$ unless some already-accepted $x$ satisfies $\alpha \cdot D(x, p') \le D(p, p')$. Slow preprocessing corresponds to calling RobustPrune with $\mathcal{V} = V \setminus \{p\}$ and $R = n-1$ for every $p \in V$.

Because the candidate list is sorted, the certifying neighbor $t$ for a non-edge $(u,v)$ also satisfies $D(u,t) \le D(u,v)$. Gollapudi et al.~\cite{sortprune} captured this extra structure in the following stronger definition.

\begin{definition}[Sorted $\alpha$-reachable graph~\cite{sortprune}]
\label{def:sortalpha}
A directed graph $G=(V,E)$ is \emph{sorted $\alpha$-reachable} with $\alpha>1$ if for every pair $u,v \in V$ and $u \neq v$, either $(u,v) \in E$ or there exists $t \in V$ satisfying all three conditions: (1) $(u,t) \in E$, (2) $D(v,t) \le D(v,u)/\alpha$, and (3) $D(u,t) \le D(u,v)$.
\end{definition}

Gollapudi et al.~\cite{sortprune} show that the DiskANN RobustPrune construction in fact uses the sorted pruning rule. Sorted $\alpha$-reachability is stronger than $\alpha$-reachability and enables tighter approximation bounds under the $\ell_2$ metric. 

DiskANN uses classical beam search (CBS) for its search procedure (Algorithm~1 of~\cite{diskann}), parameterized by a list size $L \ge k$: the algorithm maintains $L$ approximate neighbors of $q$ and returns the top $k$ at termination. Al-Jazzazi et al.~\cite{adaptivebeam} defined a new search variant called adaptive beam search (ABS) that differs from CBS in the choice of \emph{termination condition}; the general skeleton of beam search is given in \autoref{alg:beamsearch}.

The two standard termination conditions are:

\textbf{Classical beam search (CBS):} The search terminates when there exist $L \ge k$ nodes $v_1, \dots, v_L \in B \setminus \{x\}$ with $D(v_i, q) \le D(x, q)$.

\textbf{Adaptive beam search (ABS)~\cite{adaptivebeam}:} The search terminates when there exist $k$ nodes $v_1, \dots, v_k \in B \setminus \{x\}$ with $(\gamma+1) \cdot D(v_i, q) \le D(x, q)$ for parameter $\gamma \ge 0$.

Setting $\gamma = 0$ recovers the classical rule, so ABS with $\gamma = 0$ is exactly CBS at width $L=k$.

\section{Theoretical guarantees of adaptive beam search}\label{sec:theory}

The proofs of \autoref{thm:main1} and \autoref{thm:main2} share one structure, which we present as a single argument. Only one step depends on the metric and on whether the graph is sorted: the bound of Sub-case~B1 (\autoref{subsec:z0}). For the $\ell_2$ metric on sorted $\alpha$-reachable graphs, this step is a semidefinite program, whose dual certificate is in Appendix~\ref{app:thm:mainproof}. \autoref{sec:tight} shows that the bounds are tight.

Throughout this section let $n = |V|$, $s \in V$ be the starting node for search in \autoref{alg:beamsearch}, let $A = \{a_1, \ldots, a_k\}$ denote the ground-truth nearest neighbors of $q$ ordered by increasing distance, and let $B^{\star} = \{b_1, \ldots, b_k\}$ denote the top-$k$ items of $B$ at termination (see \autoref{alg:beamsearch}), also ordered by increasing distance. We assume that the distances $D(u, q)$, $u \in V$, are pairwise distinct, so that these orders are strict.

Let $B_u \subseteq B$ be the unexpanded set and $B_e = B \setminus B_u$ the expanded set. As each expansion inserts every out-neighbor into $B$, every $w \in \nout{u}$ of an expanded node $u \in B_e$ lies in $B$. The $\alpha$ and $\gamma$ parameter ranges in both theorems imply $C \ge 1$, so the statements are non-trivial.

Adaptive beam search (\autoref{alg:beamsearch}) terminates in exactly one of two ways: either (i) $B_u$ empties (line~5), or (ii) the termination condition fires (line~7). We treat the two cases separately.

\subsection{Case A: \texorpdfstring{$B_u$}{B\_u} empties}
\label{sec:caseA}
The argument for this case follows~\cite{adaptivebeam}; we give it in full. If the search terminates because $B_u$ is empty, the termination condition never held, so every popped node had all of its out-neighbors inserted into $B$. In particular, $B$ contains every node reachable from $s$ in $G$. We claim that \emph{every} node of $V$ is reachable from $s$.

Fix an arbitrary target node $u \in V$, and order the remaining nodes $x_1, x_2, \dots, x_{n-1}$ in increasing distance from $u$. For every $t \in V \setminus \{u, x_1\}$, because $\alpha > 1$ we have $\alpha \cdot D(u, t) > D(u, t) \ge D(u, x_1)$, so no other node can certify $\alpha$-reachability from $x_1$ to $u$; hence by Definition~\ref{def:alpha} there must be an edge from $x_1$ to $u$, and $u$ is reachable from $x_1$.

Suppose inductively that $u$ is reachable from each of $x_1, \dots, x_{i-1}$ for some $i > 1$. For every $j > i$ we have $\alpha \cdot D(u, x_j) > D(u, x_j) \ge D(u, x_i)$, so any certifying neighbor for the pair $(x_i, u)$ must lie in $\{u, x_1, \dots, x_{i-1}\}$; $\alpha$-reachability thus guarantees an edge from $x_i$ into this set, and so $u$ is reachable from $x_i$.

By induction $u$ is reachable from every $x_j$, in particular from $s$. Since $u$ was arbitrary, every node in $V$ is reachable from $s$, so $B = V$ at termination and ABS returns the exact top-$k$ nearest neighbors of $q$. Both theorems therefore hold in this case.

\subsection{Case B: the termination condition fires}
\label{sec:caseB}

If $B^{\star} = A$ there is nothing to prove, so pick the smallest index $i$ with $a_i \notin B^{\star}$. For every $j < i$ we have $b_j = a_j$. Since $a_i$ is the point closest to $q$ outside $\{a_1, \dots, a_{i-1}\}$ and distances are distinct, $D(a_i, q) < D(b_k, q)$; hence $a_i \notin B$, for otherwise $a_i$ would be among the top-$k$ items of $B$.

Let $x$ be the top element of $B_u$ when the termination condition fires, so that $D(x, q) \le D(u, q)$ for every $u \in B_u$. The condition provides $k$ nodes of $B \setminus \{x\}$ whose distances to $q$ are at most $D(x, q)/(\gamma+1)$. Hence $x \notin B^{\star}$, and
\[
(\gamma+1) \cdot D(b, q) \le D(x, q) \quad \text{for every } b \in B^{\star}.
\]
Now let $v \in B^{\star}$ be arbitrary. Then $D(v, q) < D(x, q) \le D(u, q)$ for every $u \in B_u$, where the first inequality is strict because $v \ne x$ and distances are distinct, so $v \notin B_u$. Since $v \in B^{\star} \subseteq B$, this forces $v \in B_e$. In particular, there is no edge from $v$ to $a_i$, since expanding $v$ would have inserted $a_i$ into $B$. Applying $\alpha$-reachability of $G$ repeatedly, starting from $v$, therefore gives nodes $t_0, t_1, \dots, t_\ell$ with $\ell \ge 2$ such that:
\begin{enumerate}[label=(\roman*), leftmargin=*, itemsep=1pt, topsep=2pt]
\item $t_0 = v$;
\item $t_\ell = a_i$;
\item $(t_z, t_{z+1}) \in E$ for $z = 0, \dots, \ell-1$;
\item $\alpha \cdot D(t_{z+1}, a_i) \le D(t_z, a_i)$ for $z = 0, \dots, \ell-1$; and,
\item when $G$ is sorted $\alpha$-reachable, additionally $D(v, t_1) \le D(v, a_i)$.
\end{enumerate}
Informally, the path $t_0 \to t_1 \to \dots \to t_\ell$ approaches $a_i$ by a factor of at least $\alpha$ per hop.

We have shown that $t_0 = v \in B_e$ and $t_\ell = a_i \notin B$. Since all out-neighbors of an expanded node are inserted in $B$, it follows that $t_{\ell-1} \notin B_e$.

Let $z^{\star}$ be the largest index with $t_{z^{\star}} \in B_e$; then $0 \le z^{\star} \le \ell - 2$. As $t_{z^{\star}} \in B_e$, we have $t_{z^{\star}+1} \in B$, and by maximality of $z^{\star}$, the node $t_{z^{\star}+1}$ is not expanded, so $t_{z^{\star}+1} \in B_u$. We consider two sub-cases.

\subsubsection{Sub-case B1: \texorpdfstring{$z^{\star} = 0$}{z* = 0}}
\label{subsec:z0}

Here $t_1 \in B_u$, so the termination condition gives $(\gamma+1) \cdot D(v, q) \le D(t_1, q)$, and $\alpha$-reachability gives $\alpha \cdot D(t_1, a_i) \le D(v, a_i)$.

\paragraph{General metric.}
Chaining the triangle inequality,
\begin{align*}
(\gamma+1) \cdot D(q, v)
&\le D(q, t_1) \\
&\le D(q, a_i) + D(a_i, t_1) \\
&\le D(q, a_i) + \frac{D(v, a_i)}{\alpha} \\
&\le D(q, a_i) + \frac{D(v, q) + D(q, a_i)}{\alpha} \\
&= \frac{D(v, q)}{\alpha} + \frac{\alpha+1}{\alpha} \cdot D(q, a_i).
\end{align*}
Rearranging,
\begin{gather*}
\left(\gamma + 1 - \frac{1}{\alpha}\right) \cdot D(q, v) \;\le\; \left(\frac{\alpha+1}{\alpha}\right) \cdot D(q, a_i)\\
\quad\Longrightarrow\quad
\frac{D(q, v)}{D(q, a_i)} \;\le\; \frac{\alpha+1}{\alpha(\gamma+1)-1}.
\end{gather*}

\paragraph{$\ell_2$ metric, sorted $\alpha$-reachable $G$.}
Sorted $\alpha$-reachability adds the constraint $D(v, t_1) \le D(v, a_i)$. The largest ratio $D(q, v)/D(q, a_i)$ compatible with the three constraints is the value of a semidefinite program over the Gram matrix of $q$, $v$, $t_1$ and $a_i$, and Appendix~\ref{app:thm:mainproof} exhibits a dual solution of value $\bigl(\alpha/(\alpha(\gamma+1)-1)\bigr)^{2}$. Hence
\[
\frac{D(q, v)}{D(q, a_i)} \;\le\; \frac{\alpha}{\alpha(\gamma+1)-1}.
\]

\subsubsection{Sub-case B2: \texorpdfstring{$z^{\star} \ge 1$}{z* >= 1}}
\label{subsec:z1}

Now $t_{z^{\star}+1} \in B_u$, so $(\gamma+1) \cdot D(v, q) \le D(t_{z^{\star}+1}, q)$. Iterating $\alpha$-reachability along the chain gives
\[
\alpha^{z^{\star}+1} \cdot D(t_{z^{\star}+1}, a_i) \;\le\; D(v, a_i),
\]
and since $z^{\star} + 1 \ge 2$, in particular $\alpha^{2} \cdot D(t_{z^{\star}+1}, a_i) \le D(v, a_i)$. This is precisely the structure of Sub-case~B1 with $\alpha$ replaced by $\alpha^{2}$ and $t_1$ by $t_{z^{\star}+1}$. The general-metric argument of Sub-case~B1 therefore gives
\[
\frac{D(q, v)}{D(q, a_i)} \;\le\; \frac{\alpha^{2}+1}{\alpha^{2}(\gamma+1)-1}.
\]

\subsection{Assembling the bound}

Combining Sub-cases~B1 and~B2, for every $v \in B^{\star}$:
\begin{equation}\label{eq:objectivevalue}
D(q, v) \;\le\; C_0 \cdot D(q, a_i),
\end{equation}
where
\[
C_0 =
\begin{cases}
\text{(general metric, $\alpha$-reachable $G$):}\\[2pt]
\displaystyle \max\!\left(\frac{\alpha+1}{\alpha(\gamma+1)-1},\frac{\alpha^{2}+1}{\alpha^{2}(\gamma+1)-1}\right) = \frac{\alpha+1}{\alpha(\gamma+1)-1},\\[12pt]
\text{($\ell_2$ metric, sorted $\alpha$-reachable $G$):}\\[2pt]
\displaystyle \max\!\left(\frac{\alpha}{\alpha(\gamma+1)-1},\frac{\alpha^{2}+1}{\alpha^{2}(\gamma+1)-1}\right).
\end{cases}
\]
The simplification in the general case follows from $\alpha > 1 \Rightarrow (\alpha+1)/(\alpha(\gamma+1)-1) \ge (\alpha^{2}+1)/(\alpha^{2}(\gamma+1)-1)$. Thus $C_0 = C$ in both theorems.

Finally, fix $1 \le j \le k$. If $j < i$ then $b_j = a_j$ and the claim is immediate. If $j \ge i$ then $b_j \in B^{\star}$, so by~\eqref{eq:objectivevalue} applied with $v = b_j$,
\[
D(b_j, q) \;\le\; C \cdot D(a_i, q) \;\le\; C \cdot D(a_j, q),
\]
where the last inequality uses that $a_1, \dots, a_k$ are sorted by increasing distance from $q$. This completes the proof for both \autoref{thm:main1} and \autoref{thm:main2}.

\subsection{Tightness}
\label{sec:tight}

The bound of \autoref{thm:main1} is attained exactly, and so is the first term of the bound of \autoref{thm:main2}. Both examples below use $k = 1$, start the search at a node $v$, and let $G$ be the complete graph on three nodes $\{v, a, t\}$ with the single edge $(v, a)$ removed, so that the search reaches the true nearest neighbor $a$ of $q$ only through $t$.

\paragraph{General metric, $\alpha$-reachable.}
Place the three nodes and the query on $\Rbb$, with $D$ the Euclidean metric on $\Rbb$ (which is also a general metric):
\[
v = \frac{\alpha+1}{\alpha(\gamma+1)-1}, \quad q = 0, \quad a = -1, \quad t = -\frac{(\alpha+1)(\gamma+1)}{\alpha(\gamma+1)-1},
\]
with $\alpha > 1$ and $0 < \gamma < 2/\alpha$. A direct check shows $\alpha \cdot D(t, a) \le D(v, a)$, so $G$ is $\alpha$-reachable. Starting from $v$, only $t$ is added to the priority queues; since $(\gamma+1) \cdot D(q, v) \le D(q, t)$, $t$ is not expanded and the search terminates with $v$ as its output. The realized ratio is
\[
\frac{D(q, v)}{D(q, a)} \;=\; \frac{\alpha+1}{\alpha(\gamma+1)-1},
\]
matching the bound of \autoref{thm:main1}.

\paragraph{$\ell_2$, sorted $\alpha$-reachable.}
Place the three nodes in $\Rbb^{2}$ together with a query:
\[
v = \left(-\tfrac{\alpha}{2},\; 0\right), \quad
t = \left(\tfrac{\alpha}{2},\; 0\right), \quad
a = \left(\tfrac{\alpha}{2} - \tfrac{1}{2\alpha},\; \sqrt{1 - \tfrac{1}{4\alpha^{2}}}\right),
\]
\[
q = \left(\frac{\alpha^{2}\bigl(1 - (\gamma+1)^{2}\bigr)}{2\bigl(\alpha(\gamma+1)^{2} + \alpha - (\gamma+1)\bigr)},\;
\frac{\alpha(\gamma+1)\sqrt{4\alpha^{2} - 1}}{2\bigl(\alpha(\gamma+1)^{2} + \alpha - (\gamma+1)\bigr)}\right),
\]
with $\alpha > 1$ and $0 < \gamma < 1/\alpha$. A direct check shows $\alpha \cdot D(t, a) \le D(v, a)$ and $D(v, t) \le D(v, a)$, so $G$ is sorted $\alpha$-reachable. Starting the search at $v$, $t$ is never expanded because $(\gamma+1) \cdot D(q, v) \le D(q, t)$, and the algorithm returns $v$. The realized ratio is
\[
\frac{D(q, v)}{D(q, a)} \;=\; \frac{\alpha}{\alpha(\gamma+1) - 1},
\]
matching the first term of the bound of \autoref{thm:main2}.

\section{Beam predictor}\label{sec:predl-main}

The guarantees of \autoref{sec:theory} bound the quality of every query's answer, but they do not say how much search each query needs, and in practice this varies widely: easy queries find their true top-$K$ after a few expansions, while hard ones need much larger lists. Beam predictor therefore gives each query its own list size $L$ for CBS, chosen to minimize the mean number of distance computations over a batch of queries subject to a target mean recall. It has three parts: an offline allocation that sets the best list size of every training query (\autoref{sec:alloc}), a model that predicts this list size from a short pilot search (\autoref{sec:predict}), and a search that continues the pilot instead of restarting (\autoref{sec:continue}).

\subsection{Cost-optimal allocation}\label{sec:alloc}

Fix a graph index and a target mean recall $r_T$, in percent. For a training query $q_i$, let $\mathrm{dist\_cmps}(q_i, L)$ and $\mathrm{recall}(q_i, L)$ be the number of distance computations and the recall@$K$, in percent, of CBS with list size $L$; one sweep of CBS over a grid $\mathcal{L}$ of list sizes measures both. The grid contains every integer in $[K, 1000]$ and every multiple of $10$ in $(1000, 10K]$. The best allocation of list sizes to the $Q$ training queries solves
\begin{equation}\label{eq:alloc}
\begin{aligned}
\min_{L_1, \dots, L_Q \in \mathcal{L}}\; & \frac{1}{Q}\sum_{i=1}^{Q} \mathrm{dist\_cmps}(q_i, L_i)\\
\text{s.t.}\; & \frac{1}{Q}\sum_{i=1}^{Q} \mathrm{recall}(q_i, L_i) \ge r_T.
\end{aligned}
\end{equation}
This is a multiple-choice knapsack problem: every query picks one list size, and the recalls must reach the target at the least total cost.

\paragraph{Reduction to recall levels.}
Only the recall level each query reaches matters for the constraint. From the sweep we therefore precompute
\begin{multline*}
\mathrm{min\_cmps}[i][r] = \min\bigl\{\mathrm{dist\_cmps}(q_i, L) : L \in \mathcal{L},\\
\mathrm{recall}(q_i, L) \ge r\bigr\}, \quad r \in \{0, 1, \dots, 100\},
\end{multline*}
with $\mathrm{min\_cmps}[i][r] = \infty$ if no $L \in \mathcal{L}$ reaches recall $r$, together with the smallest list size $\mathrm{min\_L}[i][r] \in \mathcal{L}$ that reaches recall $r$. Instead of Problem~\eqref{eq:alloc}, we work with
\begin{equation}\label{eq:reduced}
\begin{aligned}
\min_{r_1, \dots, r_Q \in \{0, \dots, 100\}}\; & \frac{1}{Q} \sum_{i=1}^{Q} \mathrm{min\_cmps}[i][r_i]\\
\text{s.t.}\; & \frac{1}{Q} \sum_{i=1}^{Q} r_i \ge r_T,
\end{aligned}
\end{equation}
and the target list size of query $q_i$ is $L^\star_i = \mathrm{min\_L}[i][r_i]$. For $K = 100$, recall@$K$ is a whole percentage, and Problem~\eqref{eq:reduced} is equivalent to Problem~\eqref{eq:alloc}. For $K = 200$ and $500$, recall@$K$ moves in steps of $0.5$ and $0.2$ points, and Problem~\eqref{eq:reduced} counts each query's recall rounded down to a whole percentage: its allocations still meet the constraint of Problem~\eqref{eq:alloc}, but they can cost more than the optimum of Problem~\eqref{eq:alloc}.

\paragraph{Lagrangian relaxation.}
We solve~\eqref{eq:reduced} approximately by putting a price $\lambda \ge 0$ on distance computations. For a fixed price the problem decouples into independent per-query problems,
\[
r_i(\lambda) = \arg\max_{r \in \{0, \dots, 100\}} \bigl(r - \lambda \cdot \mathrm{min\_cmps}[i][r]\bigr).
\]
A higher price lowers the recall each query buys, so we binary-search $\lambda$ over $[0, 5 \cdot 10^{4}]$ for $80$ steps and keep the allocation of the largest price whose mean recall still meets $r_T$. This takes $O(80 \cdot 101 \cdot Q)$ operations: less than two seconds for the $5{,}000$, $5{,}000$ and $14{,}345$ training queries of SIFT100M, DEEP100M and MSSpaceV100M in our NumPy implementation. Because the per-query cost curves need not be convex, the resulting allocation can be suboptimal.

\paragraph{Exact dynamic program.}
To measure how far the Lagrangian allocation is from optimal, we solve~\eqref{eq:reduced} exactly by dynamic programming. Let $\mathrm{dp}_i[s]$ be the minimum total cost of queries $q_1, \dots, q_i$ when their recall levels sum to $s$. Then $\mathrm{dp}_0[0] = 0$,
\[
\mathrm{dp}_i[s] = \min_{r \in \{0, \dots, 100\}} \bigl(\mathrm{dp}_{i-1}[s - r] + \mathrm{min\_cmps}[i][r]\bigr),
\]
and the optimum of~\eqref{eq:reduced} is $\min_{s \ge Q r_T} \mathrm{dp}_Q[s] / Q$. Since $s$ ranges up to $100Q$, the DP takes $O(101 \cdot 100\, Q^{2})$ time. We compare the two on $10$ random subsets of $100$ queries per in-memory setting, small enough for the DP (\autoref{tab:predl-settings}). Averaged over the subsets, the Lagrangian's allocation costs $0.07$--$0.52\%$ more distance computations than the optimum of~\eqref{eq:reduced}, and $2.1\%$ more on the worst single subset. At $K = 200$ and $500$, where the two problems differ, we also solve Problem~\eqref{eq:alloc} itself exactly, with the same DP over the number of true neighbors that each query finds instead of whole percentages: on the same subsets, the optimum of~\eqref{eq:reduced} costs $0.6$--$4.8\%$ more than that of~\eqref{eq:alloc}, and the Lagrangian's allocation costs $0.6$--$4.9\%$ more ($6.3\%$ on the worst subset). The DP took $0.12$--$0.20$\,s per $100$-query subset and $14$--$17$\,s for $1{,}000$ queries, against $3$--$6$\,ms for the Lagrangian on $100$ queries.

\subsection{Prediction from a pilot search}\label{sec:predict}

The targets $L^\star_i$ require the recall of every list size, which is known only for training queries. At query time, beam predictor first runs a pilot search, CBS with list size $L = K$ run to convergence, and summarizes it in six features. Let $d_1 \le \dots \le d_K$ be the distances of the pilot's top-$K$ results, and let $\sigma_0, \sigma_1, \dots, \sigma_T$ be the sum of the current top-$K$ distances recorded after each of the $T$ expansion steps of the pilot.

\begin{itemize}[leftmargin=*, itemsep=1pt, topsep=2pt]
\item \textbf{Scale features:} $d_1$, $d_K$, and the mean and standard deviation of $d_1, \dots, d_K$.
\item \textbf{Trajectory features:} $\mathrm{sdp} = 100 \cdot (\sigma_0 - \sigma_T)/\sigma_0$, the percentage drop of the top-$K$ distance sum, and $\mathrm{c90}$, the number of distance computations after which $\sigma_t$ first falls below $\sigma_0 - 0.9\,(\sigma_0 - \sigma_T)$.
\end{itemize}
The scale features describe the distances of the query's current neighbors; $d_K$ is the quantity in ABS's stopping rule. The trajectory features describe how quickly the sum of these distances fell during the pilot, which indicates how close the search is to convergence.

\paragraph{Model and calibration.}
We fit $\log L^\star$ by linear regression on the six features $x$, weighting query $i$ by $(L^\star_i / K)^{2}$ so that queries with large targets, which dominate the cost, are fit more closely:
\[
\widehat{\log L} \;=\; w^{\top} x + b.
\]
A single scale factor $s$ calibrates the predictions,
\[
\widehat L = \min\Bigl(10K,\ \max\bigl(K,\ \mathrm{round}\bigl(s \cdot \exp(w^{\top} x + b)\bigr)\bigr)\Bigr),
\]
where $s$ is the smallest value in $\{0.300, 0.305, 0.310, \dots\}$ for which the training queries reach mean recall $r_T$. The model, its calibration and the Lagrangian targets use only the training queries.

\subsection{Continuing the pilot}\label{sec:continue}

We implement beam predictor inside DiskANN's in-memory and SSD search loops. A trained model is a $60$-byte file holding $K$, the pilot's list size, the clipping range, $s$, $b$ and $w$. As in DiskANN, each query runs on a single thread and different threads serve different queries, so the per-query list sizes need no coordination between threads. The pilot and the main search are one execution of the search loop, in three phases.

\paragraph{Pilot.}
The search starts as CBS with list size $K$, so its sorted candidate list always holds the $K$ best candidates found so far. In addition, every newly discovered candidate is appended with its distance to a temporary per-query buffer, so the candidates that the list evicts, or never admits, are kept. After every expansion step (for in-memory search, one node; on the SSD, one batch of up to $W = 4$ reads) the search records the sum of the current top-$K$ distances and the number of distance computations so far, the trajectory behind $\mathrm{sdp}$ and $\mathrm{c90}$.

\paragraph{Prediction.}
The pilot has converged when no candidate in the list is unexpanded. The search then reads $d_1, \dots, d_K$ off the list (on the SSD as well, the distances are already in the list, so no extra I/O is needed), computes the six features in $O(K + T)$ time and evaluates $\widehat L$: six multiply-adds, one exponential and the clipping.

\paragraph{Extension.}
The search re-inserts the best $\widehat L - K$ buffered candidates that are not in the list and continues with list size $\widehat L$. For example, with $K = 100$ and $\widehat L = 420$, the $320$ best candidates outside the list are appended to it and the search continues with list size $420$; if $\widehat L = K$, the pilot's result is returned. Since the list holds the $c$ best candidates discovered so far ($c = K$ unless fewer were found), which are also the $c$ smallest entries of the buffer, the next $t = \min(\widehat L - c,\, |\mathrm{buffer}| - c)$ are found without sorting the buffer: \texttt{std::nth\_element} at position $c$ moves the list's entries to the front, a second call at position $c + t$ moves the next $t$ candidates into positions $c$ to $c + t - 1$, and only those $t$ entries are sorted. They are appended to the list as unexpanded, its capacity is raised to $\widehat L$, and the loop resumes with the same visited set until all $\widehat L$ entries are expanded. The overhead is one append per distance computation during the pilot and $O(|\mathrm{buffer}| + t \log t)$ work at the transition.

\paragraph{Alternative without a buffer.}
We also implemented a variant in which the pilot's list keeps up to $10K$ candidates in sorted order, the pilot stops once the first $K$ are expanded, and after the prediction the list is truncated to $\widehat L$ before the search continues. Its recall and distance computations were essentially the same as the buffered implementation's, but it was slower, because its list holds up to $10K$ entries for the whole pilot and every insertion into a sorted array moves all entries behind the insertion point. All our experiments use the buffered implementation.

\paragraph{Exactness of the continuation.}
For in-memory search, the continuation reproduces the CBS search at $\widehat L$. In two settings that we checked query by query (SIFT100M, sorted graph, $K = 100$ and $500$, $r = 95\%$), every query whose $\widehat L$ lies on the sweep grid reached the same recall and used exactly the same number of distance computations as in the sweep. The mean cost estimated from the sweep differs from the measured one by at most $0.20\%$, mainly because, above $L = 1000$, the sweep advances in steps of $10$ and the estimate rounds $\widehat L$ up. The recall reported in the tables differs from the estimate by at most $0.06$ points, owing to the same rounding and to DiskANN also counting as correct the results tied in distance with the $K$-th neighbor, whereas the per-query tables count only the first $K$ ground-truth ids.

\section{Experiments}\label{sec:experiments}

We now evaluate beam predictor against CBS and ABS. After the experimental setup (\autoref{sec:exp-setup}), the evaluation has three parts. First, we check whether recall moves with the relaxation parameter $\gamma$ and the pruning order in the direction that the bounds of \autoref{thm:main1} and \autoref{thm:main2} suggest, although our graphs are not known to satisfy their hypotheses (\autoref{sec:exp-theory}). Second, we compare beam predictor with CBS and ABS at equal recall (\autoref{sec:exp-compare}): query latency and SSD reads on an SSD-resident index, distance computations for in-memory search, and how the comparison changes as $K$ grows. Third, we analyze beam predictor itself (\autoref{sec:exp-analysis}): how closely its predicted list sizes follow the optimal ones, how well its calibration carries over to unseen queries, how much each group of features contributes to its savings, and how the Lagrangian targets compare with the exact optimum.

\subsection{Setup}\label{sec:exp-setup}

\paragraph{Datasets.}
We use three benchmarks from the NeurIPS'21 Billion-Scale ANN Challenge~\cite{bigann-benchmarks} (\autoref{tab:datasets}): SIFT100M, DEEP100M and MSSpaceV100M, $100$-million-point slices of SIFT1B, DEEP1B and SpaceV1B. The MSSpaceV query set contains $626$ exact duplicates among its $29{,}316$ queries; we remove them, leaving $28{,}690$ unique queries. Each query set is split once at random (seed $42$) into two halves, of $5{,}000$ queries each for SIFT100M and DEEP100M and $14{,}345$ each for MSSpaceV100M: beam predictor is trained on the first, and all results are reported on the second. The analysis in \autoref{sec:exp-analysis} repeats the split $10$ times.

\begin{table}[!htbp]
\centering
\small
\caption{NeurIPS'21 Big-ANN datasets used in our experiments.}
\label{tab:datasets}
\begin{tabular}{lrrlrc}
\toprule
Dataset & Points & Dim & Type & Queries & Metric \\
\midrule
SIFT100M        & $100{,}000{,}000$ & $128$  & uint8   & $10{,}000$ & $\ell_2$ \\
DEEP100M        & $100{,}000{,}000$ & $96$   & float32 & $10{,}000$ & $\ell_2$ \\
MSSpaceV100M      & $100{,}000{,}000$ & $100$  & int8    & $28{,}690$ & $\ell_2$ \\
\bottomrule
\end{tabular}
\end{table}

\paragraph{Hardware.}
All experiments ran on a server with two Intel Xeon Platinum $8160$ CPUs ($48$ cores, $96$ threads), $512$\,GB of RAM and a $3.2$\,TB Dell PM1725a NVMe SSD that stores the SSD indices.

\paragraph{Indices.}
All graphs are built with DiskANN's Vamana algorithm~\cite{diskann}, i.e., with fast preprocessing (\autoref{sec:prelim}), with maximum out-degree $R = 32$ and build list size $L_{\mathrm{build}} = 100$. DiskANN's pruning parameter is set to $\alpha = 1.2$. DiskANN sorts the candidate list before pruning, the rule that makes slow-preprocessing graphs sorted $\alpha$-reachable (\autoref{sec:prelim}); we call the graph it builds the \emph{sorted} graph and evaluate search on it, in memory and on the SSD. Like other fast-preprocessing graphs, it is not known to be $\alpha$-reachable, so our guarantees do not apply to it. The SSD index stores each node's vector and neighbor list on the SSD; we build it by writing the in-memory graph in DiskANN's SSD layout, so in-memory and SSD search traverse the same graph. For the theory validation (\autoref{sec:exp-theory}) we also build, with the same parameters, a \emph{shuffled} graph, which permutes the candidate list at random before pruning; with slow preprocessing, this rule yields graphs that are $\alpha$-reachable but not necessarily sorted $\alpha$-reachable. As in DiskANN, the SSD search computes the distances of the neighbors of each node it expands from compressed codes kept in RAM, one byte per dimension, and ranks its results by full-precision distance; the codes are nearly lossless: on a sample of $100{,}000$ vectors, their relative reconstruction error (the Frobenius norm of the error divided by that of the vectors' deviation from their mean) is below $1\%$ on all three datasets. On the SSD index, the cost $\mathrm{dist\_cmps}$ in Problem~\eqref{eq:alloc} is DiskANN's per-query counter of distance computations, which adds the number of neighbors of every expanded node and adds one more for each newly discovered neighbor of a node read from the SSD; it is thus a proxy for the work of the search rather than an exact count of distance computations. SSD searches use $W = 4$~\cite{diskann-github}, i.e., they read up to $4$ nodes from the SSD per step; they keep $10{,}000$ nodes cached in RAM, read the SSD with direct I/O (bypassing the page cache) and run on a single thread, and every operating point is timed $5$ times.

\paragraph{Search variants.}
We compare three procedures. \textbf{CBS} uses the same list size $L$ for every query, \textbf{ABS}~\cite{adaptivebeam} the same relaxation parameter $\gamma$, and \textbf{beam predictor} (\autoref{sec:predl-main}) a list size predicted for each query by a model trained on the training half.

\paragraph{Beam predictor configuration.}
\autoref{tab:predl-setup} lists, for each $K$, the list size of the pilot search, the grid $\mathcal{L}$ of list sizes swept on the training queries to build the per-query tables of \autoref{sec:alloc}, the range to which predictions are clipped, and the target recalls. The grid has $901$ values for every $K$: step $1$ up to $1000$ and step $10$ above. One sweep serves every $K$ of a dataset. The traversal of CBS depends only on $L$; $K$ only decides how many of the final, sorted results are returned. We therefore search each $L$ in the union of the grids of \autoref{tab:predl-setup} once, return the top $K'$ results, where $K'$ is the largest studied $K$ not exceeding $L$, and read recall@$K$ for every $K \le K'$ off the same result list; we checked that this reproduces separate per-$K$ sweeps exactly. On the SSD index the sweep runs the same search code with the index file in the page cache and many threads; its outputs are identical to direct-I/O runs, and only the timed runs use direct I/O on a single thread. This sweep is the main offline cost of beam predictor: run once for all three $K$ and all queries, although training uses only the training half, it took $1.7$\,h for SIFT100M, $1.8$\,h for DEEP100M and $3.4$\,h for MSSpaceV100M on our server. Computing the targets and fitting and calibrating the model then take less than two seconds per setting. Neither cost is included in the latencies we report, which do include the pilot search, the computation of the features and the prediction.

\begin{table}[!htbp]
\centering
\small
\caption{Beam predictor configuration for each $K$. Pilot: list size of the pilot search; $\mathcal{L}$: list sizes swept on the training queries; predictions are clipped to $[K, 10K]$.}
\label{tab:predl-setup}
\setlength{\tabcolsep}{4pt}
\begin{tabular}{@{}rrlrl@{}}
\toprule
$K$ & Pilot & Sweep grid $\mathcal{L}$ & $|\mathcal{L}|$ & $r$ (\%) \\
\midrule
$100$ & $100$ & $100, 101, \dots, 1000$ & $901$ & $90, 95$ \\
$200$ & $200$ & $200, \dots, 1000, 1010, \dots, 2000$ & $901$ & $90, 95$ \\
$500$ & $500$ & $500, \dots, 1000, 1010, \dots, 5000$ & $901$ & $95$ \\
\bottomrule
\end{tabular}
\end{table}

\paragraph{Other per-query methods.}
We do not compare beam predictor experimentally with the per-query methods of \autoref{sec:related}, so we make no claim about how it performs relative to them. Their objectives differ from ours: several aim for every query to reach a declared recall target~\cite{darth,adaef,annie}, whereas beam predictor minimizes the total cost of a batch of queries at a target mean recall. They were also designed and evaluated for in-memory indices such as HNSW or partition-based (IVF) indices, whereas our evaluation centers on DiskANN graphs and their SSD search. A comparison under a common objective is left for future work.

\paragraph{Protocol.}
For each dataset, number of neighbors $K$ (the $k$ of \autoref{sec:prelim}) and target mean recall $r$, we train beam predictor on the training half and measure its mean recall@$K$ on the test half. We then set the $L$ of CBS and the $\gamma$ of ABS so that their test recall matches the predictor's to within $0.05$ percentage points, and compare cost at that recall. To find these values, we evaluate CBS on the test half at $L \in \{1.2, 1.6, 2, 2.6, 3.2, 4, 5, 6.5, 8, 10\} \cdot K$ and ABS at $\gamma \in \{0.02, 0.04, \dots, 0.2\}$, then repeatedly subdivide the interval that brackets the predictor's test recall, extending the range when needed, until a value is within $0.05$ percentage points of it; $L$ is an integer and $\gamma$ has four decimals. The baselines are thus tuned on the test queries themselves. We use $r \in \{90, 95\}\%$ for $K \in \{100, 200\}$ and $r = 95\%$ for $K = 500$; at $K = 500$, CBS at its smallest list size $L = K$ already exceeds $90\%$ recall on SIFT100M. On the SSD we report mean query latency and mean SSD reads (I/Os) per query; for in-memory search we report mean distance computations per query, one per newly discovered node.

\subsection{Theory validation: \texorpdfstring{$\gamma$}{gamma} and the pruning order}\label{sec:exp-theory}

\begin{figure}[!htbp]
  \centering
  \includegraphics[width=\textwidth]{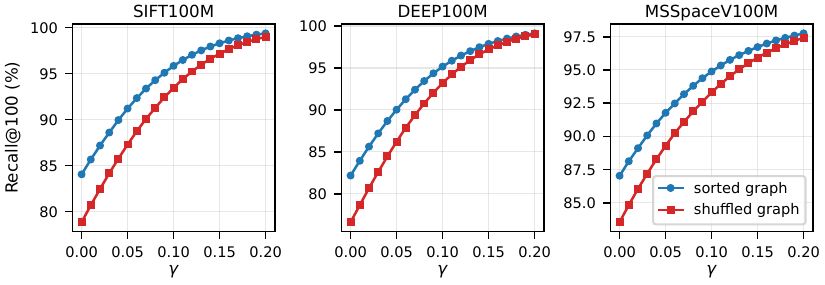}
  \caption{Recall@$100$ of ABS versus the relaxation parameter~$\gamma$ on the three datasets, for the sorted graph and the shuffled graph (\autoref{sec:exp-setup}), both built with DiskANN ($R{=}32$, $L_{\text{build}}{=}100$), over all queries of each dataset.}
  \label{fig:abs-sort-unsort}
\end{figure}

\autoref{thm:main1} and \autoref{thm:main2} bound the worst-case approximation ratio on $\alpha$-reachable and sorted $\alpha$-reachable graphs: the bound shrinks monotonically in $\gamma$ and is smaller for sorted $\alpha$-reachable graphs under $\ell_2$. Our graphs are built with fast preprocessing and are not known to satisfy either definition, and we measure recall rather than the worst-case ratio, so this experiment does not test the theorems; it only checks whether recall moves in the direction the bounds suggest. We ran ABS with $\gamma \in \{0, 0.01, \dots, 0.2\}$ and $K = 100$ on the sorted and the shuffled graph and measured recall@$100$. \autoref{fig:abs-sort-unsort} shows that on all three datasets recall@$100$ rises monotonically with $\gamma$ on both graphs, and that the sorted graph reaches a higher recall than the shuffled graph at every $\gamma$.

\subsection{Comparison with CBS and ABS}\label{sec:exp-compare}

We compare beam predictor with CBS and ABS on the SSD index and for in-memory search, both on the sorted graph.

\begin{table}[!htbp]
\centering
\small
\caption{SSD index, sorted graph: beam predictor (BP) versus CBS and ABS in mean query latency in ms (mean {\scriptsize$\pm$ std.\ dev.} over $5$ runs) and mean SSD reads (I/Os) per query at matched mean recall@$K$ on the test half. $r$ is the target recall (\%); the $L$ of CBS and the $\gamma$ of ABS are chosen so that their recall matches that of beam predictor within $0.05$ percentage points. The last six columns give relative changes (\%) in latency (Lat.) and I/Os: $\Delta_{\mathrm{CBS}} = (\mathrm{BP} - \mathrm{CBS})/\mathrm{CBS}$, $\Delta_{\mathrm{ABS}} = (\mathrm{BP} - \mathrm{ABS})/\mathrm{ABS}$ and $\Delta_{\mathrm{ABS/CBS}} = (\mathrm{ABS} - \mathrm{CBS})/\mathrm{CBS}$; negative values mean that BP, or ABS for $\Delta_{\mathrm{ABS/CBS}}$, is cheaper. Bold marks the lowest latency and the lowest I/Os in each row. Single search thread, $W = 4$, direct I/O.}
\label{tab:ssd}
\setlength{\tabcolsep}{2.5pt}
\fitwidth{\begin{tabular}{@{}rrrrrrrrrrrrrr@{}}
\toprule
 & & \multicolumn{3}{c}{Latency (ms)} & \multicolumn{3}{c}{I/Os per query} & \multicolumn{2}{c}{$\Delta_{\mathrm{CBS}}$ (\%)} & \multicolumn{2}{c}{$\Delta_{\mathrm{ABS}}$ (\%)} & \multicolumn{2}{c}{$\Delta_{\mathrm{ABS/CBS}}$ (\%)} \\
\cmidrule(lr){3-5} \cmidrule(lr){6-8} \cmidrule(lr){9-10} \cmidrule(lr){11-12} \cmidrule(l){13-14}
$K$ & $r$ & CBS & ABS & BP & CBS & ABS & BP & Lat. & I/Os & Lat. & I/Os & Lat. & I/Os \\
\midrule
\multicolumn{14}{@{}l}{\textit{SIFT100M}} \\
100 & 90 & $8.42${\scriptsize$\pm0.03$} & $8.06${\scriptsize$\pm0.09$} & $\mathbf{8.00}${\scriptsize$\pm0.03$} & $170.5$ & $163.0$ & $\mathbf{159.6}$ & $-4.9$ & $-6.4$ & $-0.8$ & $-2.1$ & $-4.2$ & $-4.4$ \\
100 & 95 & $14.78${\scriptsize$\pm0.02$} & $13.24${\scriptsize$\pm0.07$} & $\mathbf{12.68}${\scriptsize$\pm0.15$} & $305.0$ & $269.9$ & $\mathbf{261.0}$ & $-14.2$ & $-14.5$ & $-4.2$ & $-3.3$ & $-10.4$ & $-11.5$ \\
200 & 90 & $\mathbf{11.37}${\scriptsize$\pm0.16$} & $11.38${\scriptsize$\pm0.19$} & $11.45${\scriptsize$\pm0.05$} & $232.7$ & $231.1$ & $\mathbf{229.2}$ & $+0.6$ & $-1.5$ & $+0.6$ & $-0.8$ & $+0.0$ & $-0.7$ \\
200 & 95 & $20.43${\scriptsize$\pm0.08$} & $18.92${\scriptsize$\pm0.16$} & $\mathbf{17.69}${\scriptsize$\pm0.10$} & $421.3$ & $387.9$ & $\mathbf{360.8}$ & $-13.4$ & $-14.4$ & $-6.5$ & $-7.0$ & $-7.4$ & $-7.9$ \\
500 & 95 & $32.10${\scriptsize$\pm0.15$} & $31.56${\scriptsize$\pm0.25$} & $\mathbf{29.79}${\scriptsize$\pm0.13$} & $667.3$ & $646.3$ & $\mathbf{609.0}$ & $-7.2$ & $-8.7$ & $-5.6$ & $-5.8$ & $-1.7$ & $-3.1$ \\
\midrule
\multicolumn{14}{@{}l}{\textit{DEEP100M}} \\
100 & 90 & $9.70${\scriptsize$\pm0.11$} & $\mathbf{9.18}${\scriptsize$\pm0.12$} & $9.21${\scriptsize$\pm0.04$} & $205.2$ & $192.0$ & $\mathbf{190.5}$ & $-5.0$ & $-7.2$ & $+0.3$ & $-0.8$ & $-5.3$ & $-6.4$ \\
100 & 95 & $18.58${\scriptsize$\pm0.12$} & $15.95${\scriptsize$\pm0.08$} & $\mathbf{15.82}${\scriptsize$\pm0.04$} & $397.5$ & $\mathbf{335.9}$ & $\mathbf{335.9}$ & $-14.8$ & $-15.5$ & $-0.8$ & $+0.0$ & $-14.2$ & $-15.5$ \\
200 & 90 & $13.25${\scriptsize$\pm0.13$} & $13.14${\scriptsize$\pm0.11$} & $\mathbf{12.90}${\scriptsize$\pm0.10$} & $280.3$ & $272.8$ & $\mathbf{268.1}$ & $-2.6$ & $-4.4$ & $-1.8$ & $-1.7$ & $-0.8$ & $-2.7$ \\
200 & 95 & $25.36${\scriptsize$\pm0.26$} & $22.62${\scriptsize$\pm0.24$} & $\mathbf{21.90}${\scriptsize$\pm0.27$} & $540.8$ & $476.5$ & $\mathbf{464.7}$ & $-13.7$ & $-14.1$ & $-3.2$ & $-2.5$ & $-10.8$ & $-11.9$ \\
500 & 95 & $39.74${\scriptsize$\pm0.32$} & $37.44${\scriptsize$\pm0.36$} & $\mathbf{35.51}${\scriptsize$\pm0.29$} & $851.7$ & $793.2$ & $\mathbf{754.0}$ & $-10.6$ & $-11.5$ & $-5.2$ & $-4.9$ & $-5.8$ & $-6.9$ \\
\midrule
\multicolumn{14}{@{}l}{\textit{MSSpaceV100M}} \\
100 & 90 & $7.46${\scriptsize$\pm0.05$} & $\mathbf{7.01}${\scriptsize$\pm0.07$} & $7.11${\scriptsize$\pm0.20$} & $159.3$ & $\mathbf{147.2}$ & $150.6$ & $-4.6$ & $-5.5$ & $+1.4$ & $+2.3$ & $-5.9$ & $-7.6$ \\
100 & 95 & $20.61${\scriptsize$\pm0.15$} & $14.77${\scriptsize$\pm0.12$} & $\mathbf{14.51}${\scriptsize$\pm0.13$} & $448.3$ & $322.1$ & $\mathbf{312.1}$ & $-29.6$ & $-30.4$ & $-1.8$ & $-3.1$ & $-28.3$ & $-28.2$ \\
200 & 90 & $11.40${\scriptsize$\pm0.13$} & $\mathbf{11.15}${\scriptsize$\pm0.16$} & $11.25${\scriptsize$\pm0.22$} & $247.3$ & $\mathbf{238.9}$ & $241.1$ & $-1.3$ & $-2.5$ & $+0.9$ & $+1.0$ & $-2.1$ & $-3.4$ \\
200 & 95 & $29.98${\scriptsize$\pm0.98$} & $23.40${\scriptsize$\pm0.45$} & $\mathbf{22.96}${\scriptsize$\pm0.13$} & $662.9$ & $515.8$ & $\mathbf{495.5}$ & $-23.4$ & $-25.2$ & $-1.9$ & $-3.9$ & $-22.0$ & $-22.2$ \\
500 & 95 & $60.21${\scriptsize$\pm0.16$} & $50.03${\scriptsize$\pm0.21$} & $\mathbf{47.71}${\scriptsize$\pm0.30$} & $1{,}309.1$ & $1{,}084.6$ & $\mathbf{1{,}039.4}$ & $-20.8$ & $-20.6$ & $-4.6$ & $-4.2$ & $-16.9$ & $-17.1$ \\
\bottomrule
\end{tabular}}
\end{table}

\paragraph{SSD results.}\label{sec:exp-ssd}
\autoref{tab:ssd} reports latency and SSD reads. Beam predictor has a lower mean latency than CBS in $14$ of the $15$ settings: by $2.6$--$29.6\%$ in $13$ of them. Its mean latency is lower than that of ABS in $11$ settings, by up to $6.5\%$, including all nine settings at $r = 95\%$. In the other four settings, all at $r = 90\%$, it is slower than ABS by at most $1.4\%$. It issues fewer SSD reads than CBS in every setting and fewer than ABS in $12$. At $r = 90\%$ the gains are small because very little search is left to reallocate: at $K = 200$, CBS with its smallest list size $L = K$ already reaches $88.9\%$ recall on both SIFT100M and MSSpaceV100M.

The latencies of beam predictor include the pilot search, the computation of the features and the prediction. Its gains are in mean latency: because for the queries that it predicts to be hard, it allocates list sizes of up to $10 \cdot K$, its $99.9$th-percentile latency, averaged over the five runs, is higher than that of CBS in $13$ of the $15$ settings, by up to a factor of $2.1$, and higher than that of ABS in $9$. ABS has the heaviest tail on MSSpaceV100M, where at $r = 95\%$ its $99.9$th-percentile latency is up to $2.3$ times that of beam predictor. \autoref{tab:tails} in Appendix~\ref{app:tails} lists these latencies, together with the median and standard deviation of per-query recall. At matched mean recall, the per-query recall of beam predictor has a lower standard deviation than that of CBS in all $15$ settings and than that of ABS in $13$, and a median at most $2$ points below that of CBS, so its per-query recall is more concentrated around the mean.

\begin{table}[!htbp]
\centering
\small
\caption{In-memory search, sorted graph: mean distance computations per query at matched mean recall@$K$ on the test half. BP: beam predictor; $r$, recall matching, the relative changes $\Delta$ (\%) and bold as in \autoref{tab:ssd}.}
\label{tab:mem-sorted}
\setlength{\tabcolsep}{3.5pt}
\begin{tabular}{@{}rrrrrrrr@{}}
\toprule
 & & \multicolumn{3}{c}{Mean dist\_cmps} & \multicolumn{3}{c}{Relative change (\%)} \\
\cmidrule(lr){3-5} \cmidrule(l){6-8}
$K$ & $r$ & CBS & ABS & BP & $\Delta_{\mathrm{CBS}}$ & $\Delta_{\mathrm{ABS}}$ & $\Delta_{\mathrm{ABS/CBS}}$ \\
\midrule
\multicolumn{8}{@{}l}{\textit{SIFT100M}} \\
100 & 90 & $4{,}031$ & $3{,}909$ & $\mathbf{3{,}839}$ & $-4.8$ & $-1.8$ & $-3.0$ \\
100 & 95 & $6{,}786$ & $6{,}237$ & $\mathbf{6{,}096}$ & $-10.2$ & $-2.3$ & $-8.1$ \\
200 & 90 & $5{,}324$ & $5{,}289$ & $\mathbf{5{,}253}$ & $-1.3$ & $-0.7$ & $-0.7$ \\
200 & 95 & $8{,}934$ & $8{,}482$ & $\mathbf{8{,}153}$ & $-8.7$ & $-3.9$ & $-5.1$ \\
500 & 95 & $13{,}462$ & $13{,}231$ & $\mathbf{12{,}734}$ & $-5.4$ & $-3.8$ & $-1.7$ \\
\midrule
\multicolumn{8}{@{}l}{\textit{DEEP100M}} \\
100 & 90 & $4{,}594$ & $4{,}390$ & $\mathbf{4{,}373}$ & $-4.8$ & $-0.4$ & $-4.4$ \\
100 & 95 & $8{,}203$ & $7{,}358$ & $\mathbf{7{,}348}$ & $-10.4$ & $-0.1$ & $-10.3$ \\
200 & 90 & $6{,}018$ & $5{,}931$ & $\mathbf{5{,}879}$ & $-2.3$ & $-0.9$ & $-1.4$ \\
200 & 95 & $10{,}668$ & $9{,}872$ & $\mathbf{9{,}792}$ & $-8.2$ & $-0.8$ & $-7.5$ \\
500 & 95 & $15{,}808$ & $15{,}262$ & $\mathbf{14{,}972}$ & $-5.3$ & $-1.9$ & $-3.5$ \\
\midrule
\multicolumn{8}{@{}l}{\textit{MSSpaceV100M}} \\
100 & 90 & $3{,}162$ & $2{,}962$ & $\mathbf{2{,}961}$ & $-6.4$ & $-0.0$ & $-6.3$ \\
100 & 95 & $7{,}777$ & $6{,}096$ & $\mathbf{5{,}937}$ & $-23.7$ & $-2.6$ & $-21.6$ \\
200 & 90 & $4{,}569$ & $\mathbf{4{,}480}$ & $\mathbf{4{,}480}$ & $-2.0$ & $-0.0$ & $-2.0$ \\
200 & 95 & $11{,}440$ & $9{,}295$ & $\mathbf{9{,}105}$ & $-20.4$ & $-2.1$ & $-18.7$ \\
500 & 95 & $21{,}134$ & $18{,}089$ & $\mathbf{17{,}983}$ & $-14.9$ & $-0.6$ & $-14.4$ \\
\bottomrule
\end{tabular}
\end{table}

\paragraph{In-memory results.}\label{sec:exp-mem}
\autoref{tab:mem-sorted} reports distance computations for in-memory search on the sorted graph. Beam predictor needs fewer distance computations than CBS in every setting, up to $23.7\%$ fewer, and no more than ABS: between $0.0\%$ and $3.9\%$ fewer.

\paragraph{Effect of $K$.}\label{sec:exp-k}
Al-Jazzazi et al.~\cite[App.~C.1]{adaptivebeam} observe that on navigable graphs ABS gains less over CBS at $K = 100$ than at $K = 1$ and $10$. The $\Delta_{\mathrm{ABS/CBS}}$ columns of \autoref{tab:ssd} and \autoref{tab:mem-sorted} show that this trend continues for $K \in \{100, 200, 500\}$: at $r = 95\%$, the saving of ABS over CBS shrinks as $K$ grows on every dataset, on the SSD and in memory. For instance, on the SSD index ABS is $10.4\%$, $7.4\%$ and $1.7\%$ faster than CBS on SIFT100M at $K = 100$, $200$ and $500$. Beam predictor keeps more of its saving: it is $14.2\%$, $13.4\%$ and $7.2\%$ faster than CBS. On the SSD, its lead over ABS at $r = 95\%$ is therefore larger at $K = 500$ than at $K = 100$ on every dataset: it grows from $0.8\%$ to $5.2\%$ on DEEP100M and from $1.8\%$ to $4.6\%$ on MSSpaceV100M, and only slightly, from $4.2\%$ to $5.6\%$, on SIFT100M. At $K = 500$ beam predictor has the lowest latency and the fewest SSD reads on all three datasets. In memory, its saving over ABS at $K = 500$ is $3.8\%$ on SIFT100M, $1.9\%$ on DEEP100M and $0.6\%$ on MSSpaceV100M.

\subsection{Analysis of beam predictor}\label{sec:exp-analysis}

\begin{table}[!htbp]
\centering
\small
\caption{Analysis of beam predictor on the in-memory sorted graph ($K = 100$), averaged over $10$ random $50/50$ splits of the queries into a training and a test half. In the cost rows, negative values mean fewer distance computations than CBS, the indented rows retrain beam predictor with fewer features, and the scale factor of every variant is tuned on the test half so that all variants are compared at the same test recall.}
\label{tab:predl-analysis}
\setlength{\tabcolsep}{4pt}
\begin{tabular*}{\textwidth}{@{\extracolsep{\fill}}lrrrrrr@{}}
\toprule
 & \multicolumn{2}{c}{SIFT100M} & \multicolumn{2}{c}{DEEP100M} & \multicolumn{2}{c}{MSSpaceV100M} \\
\cmidrule(lr){2-3} \cmidrule(lr){4-5} \cmidrule(l){6-7}
Target recall $r$ (\%) & $90$ & $95$ & $90$ & $95$ & $90$ & $95$ \\
\midrule
\multicolumn{7}{@{}l}{\textit{Spearman's $\rho$ of $\widehat L$ vs.\ $L^\star$}} \\
training half & $0.63$ & $0.78$ & $0.67$ & $0.77$ & $0.41$ & $0.74$ \\
test half & $0.63$ & $0.78$ & $0.66$ & $0.77$ & $0.40$ & $0.74$ \\
\midrule
\multicolumn{7}{@{}l}{\textit{Calibration: test recall (\%)}} \\
mean over splits & $90.00$ & $94.99$ & $89.97$ & $94.94$ & $90.02$ & $95.02$ \\
worst split, $|\mathrm{recall} - r|$ & $0.41$ & $0.21$ & $0.38$ & $0.26$ & $0.45$ & $0.22$ \\
\midrule
\multicolumn{7}{@{}l}{\textit{Cost vs.\ CBS at equal recall (\%)}} \\
\textbf{beam predictor} & $\mathbf{-4.8}$ & $\mathbf{-10.0}$ & $\mathbf{-4.7}$ & $\mathbf{-10.3}$ & $\mathbf{-6.8}$ & $\mathbf{-23.7}$ \\
\quad no trajectory features & $-4.5$ & $-9.4$ & $-4.3$ & $-9.1$ & $-6.5$ & $-22.1$ \\
\quad trajectory features only & $-4.1$ & $-8.6$ & $-3.3$ & $-7.5$ & $-5.8$ & $-21.2$ \\
\quad $d_K$ only & $-3.1$ & $-6.4$ & $-1.9$ & $-4.6$ & $-4.5$ & $-14.6$ \\
\midrule
\multicolumn{7}{@{}l}{\textit{Lagrangian vs.\ exact DP: excess cost (\%)}} \\
$100$-query subsets & $+0.20$ & $+0.24$ & $+0.31$ & $+0.18$ & $+0.45$ & $+0.30$ \\
\bottomrule
\end{tabular*}
\end{table}

We analyze the predictor on the in-memory sorted graph with $K = 100$, over $10$ random $50/50$ splits of the queries (\autoref{tab:predl-analysis}), and then at $K = 200$ and $500$. To compare many variants we compute each one's cost from the per-query CBS sweep instead of running the search; for beam predictor itself, the cost measured online matched this estimate to within $0.20\%$ in every in-memory setting (\autoref{sec:continue}).

\paragraph{Prediction accuracy.}
On the training half, the predicted list sizes rank the queries similarly to the Lagrangian-optimal $L^\star$: Spearman's rank correlation is $0.63$ and $0.78$ on SIFT100M, $0.67$ and $0.77$ on DEEP100M, and $0.41$ and $0.74$ on MSSpaceV100M, at $r = 90\%$ and $95\%$. On the test half, the correlation is within $0.003$ of these values in every setting, so the linear model does not overfit.

\paragraph{Calibration.}
With the scale factor tuned on the training half, the test recall averaged over the $10$ splits is within $0.06$ percentage points of $r$ in all six settings, and within $0.45$ points on every individual split.

\paragraph{Components.}
The cost rows of \autoref{tab:predl-analysis} give the cost of each variant relative to CBS at the same test recall; to compare the variants at equal recall, the scale factor of each one, including the full beam predictor, is tuned on the test half rather than the training half. The trajectory features add $0.3$--$1.6$ points of saving over the distance features alone, and each group alone recovers most of the saving. The largest top-$K$ distance $d_K$, the quantity in ABS's stopping rule, recovers $40$--$66\%$ of it on its own, the least of all variants.

\paragraph{Lagrangian versus exact optimization.}
The allocation problem that defines the targets can be solved exactly by dynamic programming, but the DP's running time grows quadratically with the number of queries (\autoref{sec:alloc}): at $K = 100$, it takes $0.12$--$0.19$\,s for $100$ queries and $14$--$16$\,s for $1{,}000$. The Lagrangian takes less than two seconds on a full training half, and on $100$-query subsets, where the DP is cheap, its allocation costs on average $0.18$--$0.45\%$ more distance computations than the optimum.

\begin{table}[!htbp]
\centering
\small
\begin{minipage}[t]{0.55\linewidth}
\centering
\caption{$t$-values of the six features in the weighted regression on $\log L^\star$ (in-memory sorted graph, mean over $10$ random $50/50$ query splits).}
\label{tab:tvalues}
\setlength{\tabcolsep}{4pt}
\begin{tabular}{@{}rrrrrrrr@{}}
\toprule
$K$ & $r$ & $d_1$ & $d_K$ & mean & std & $\mathrm{sdp}$ & $\mathrm{c90}$ \\
\midrule
\multicolumn{8}{@{}l}{\textit{SIFT100M}} \\
100 & 90 & $2.8$ & $-12.4$ & $13.4$ & $2.8$ & $-10.8$ & $6.0$ \\
100 & 95 & $1.0$ & $-15.2$ & $16.8$ & $2.2$ & $-16.6$ & $6.2$ \\
200 & 90 & $2.9$ & $-7.0$ & $7.5$ & $1.6$ & $-3.4$ & $6.5$ \\
200 & 95 & $2.9$ & $-15.3$ & $16.6$ & $3.9$ & $-9.9$ & $19.4$ \\
500 & 95 & $0.7$ & $-8.2$ & $8.9$ & $1.9$ & $4.4$ & $34.1$ \\
\midrule
\multicolumn{8}{@{}l}{\textit{DEEP100M}} \\
100 & 90 & $5.4$ & $-17.9$ & $18.9$ & $7.3$ & $-10.2$ & $5.2$ \\
100 & 95 & $4.9$ & $-22.6$ & $24.1$ & $9.5$ & $-16.0$ & $7.0$ \\
200 & 90 & $4.3$ & $-14.1$ & $14.9$ & $5.8$ & $2.9$ & $16.2$ \\
200 & 95 & $3.5$ & $-24.3$ & $26.0$ & $9.4$ & $-2.6$ & $18.1$ \\
500 & 95 & $0.2$ & $-13.5$ & $15.0$ & $2.5$ & $25.9$ & $40.4$ \\
\midrule
\multicolumn{8}{@{}l}{\textit{MSSpaceV100M}} \\
100 & 90 & $4.8$ & $-0.8$ & $1.8$ & $-3.2$ & $-17.0$ & $-12.1$ \\
100 & 95 & $6.7$ & $-23.9$ & $26.6$ & $8.3$ & $-40.8$ & $-2.4$ \\
200 & 90 & $5.1$ & $1.9$ & $-1.0$ & $-4.2$ & $-10.8$ & $-5.6$ \\
200 & 95 & $10.4$ & $-29.1$ & $31.4$ & $12.6$ & $-47.6$ & $8.2$ \\
500 & 95 & $10.4$ & $-37.4$ & $40.9$ & $16.4$ & $-27.0$ & $37.1$ \\
\bottomrule
\end{tabular}
\end{minipage}\hfill
\begin{minipage}[t]{0.41\linewidth}
\centering
\caption{Mean distance computations relative to CBS at the same test recall (\%; scale factor of each variant tuned on the test half), in-memory sorted graph, mean over $10$ random $50/50$ query splits. BP: beam predictor; no traj.: without the trajectory features; traj.: trajectory features only; $d_K$: $d_K$ only.}
\label{tab:components}
\setlength{\tabcolsep}{4pt}
\begin{tabular}{@{}rrrrrr@{}}
\toprule
$K$ & $r$ & BP & no traj. & traj. & $d_K$ \\
\midrule
\multicolumn{6}{@{}l}{\textit{SIFT100M}} \\
100 & 90 & $-4.8$ & $-4.5$ & $-4.1$ & $-3.1$ \\
100 & 95 & $-10.0$ & $-9.4$ & $-8.6$ & $-6.4$ \\
200 & 90 & $-1.4$ & $-1.3$ & $-1.0$ & $-0.7$ \\
200 & 95 & $-8.5$ & $-8.2$ & $-7.3$ & $-5.2$ \\
500 & 95 & $-5.2$ & $-4.7$ & $-4.8$ & $-2.7$ \\
\midrule
\multicolumn{6}{@{}l}{\textit{DEEP100M}} \\
100 & 90 & $-4.7$ & $-4.3$ & $-3.3$ & $-1.9$ \\
100 & 95 & $-10.3$ & $-9.1$ & $-7.5$ & $-4.6$ \\
200 & 90 & $-2.5$ & $-2.3$ & $-1.8$ & $-1.0$ \\
200 & 95 & $-8.9$ & $-8.1$ & $-6.6$ & $-4.3$ \\
500 & 95 & $-5.7$ & $-5.1$ & $-4.5$ & $-2.3$ \\
\midrule
\multicolumn{6}{@{}l}{\textit{MSSpaceV100M}} \\
100 & 90 & $-6.8$ & $-6.5$ & $-5.8$ & $-4.5$ \\
100 & 95 & $-23.7$ & $-22.1$ & $-21.2$ & $-14.6$ \\
200 & 90 & $-2.3$ & $-2.1$ & $-1.8$ & $-1.4$ \\
200 & 95 & $-19.3$ & $-17.3$ & $-16.8$ & $-9.9$ \\
500 & 95 & $-15.0$ & $-13.7$ & $-12.1$ & $-6.2$ \\
\bottomrule
\end{tabular}
\end{minipage}
\end{table}

\paragraph{Feature significance.}
\autoref{tab:tvalues} gives the $t$-value of each feature in the weighted regression at every $K$ and $r$. The trajectory feature $\mathrm{c90}$ gains weight as $K$ grows: for example, at $r = 95\%$, its $t$-value rises from $6.2$ to $19.4$ to $34.1$ on SIFT100M. Because the features are correlated, a $t$-value measures a feature's contribution given the others; \autoref{tab:components} measures the effect of removing feature groups on cost.

\begin{table}[!htbp]
\centering
\small
\caption{In-memory search, sorted graph: beam predictor in every setting of \autoref{tab:mem-sorted} (mean over $10$ random $50/50$ query splits). $\rho$: Spearman correlation between predicted and Lagrangian-optimal list sizes on the test half; Recall: mean test recall (\%); Worst: largest $|\mathrm{recall} - r|$ on any one split; Lagr.\ vs DP: excess cost of the Lagrangian over the optimum of Problem~\eqref{eq:reduced}, computed exactly by DP, on $100$-query subsets (\%; mean, and largest subset).}
\label{tab:predl-settings}
\setlength{\tabcolsep}{4pt}
\begin{tabular}{@{}rrrrrl@{}}
\toprule
$K$ & $r$ & $\rho$ & Recall & Worst & Lagr.\ vs DP \\
\midrule
\multicolumn{6}{@{}l}{\textit{SIFT100M}} \\
100 & 90 & $0.63$ & $90.00$ & $0.41$ & $+0.20$ (max $0.7$) \\
100 & 95 & $0.78$ & $94.99$ & $0.21$ & $+0.24$ (max $1.0$) \\
200 & 90 & $0.47$ & $89.98$ & $0.47$ & $+0.07$ (max $0.2$) \\
200 & 95 & $0.80$ & $94.97$ & $0.23$ & $+0.11$ (max $0.3$) \\
500 & 95 & $0.74$ & $95.00$ & $0.24$ & $+0.16$ (max $0.3$) \\
\midrule
\multicolumn{6}{@{}l}{\textit{DEEP100M}} \\
100 & 90 & $0.66$ & $89.97$ & $0.38$ & $+0.31$ (max $1.3$) \\
100 & 95 & $0.77$ & $94.94$ & $0.26$ & $+0.18$ (max $0.5$) \\
200 & 90 & $0.60$ & $89.94$ & $0.33$ & $+0.19$ (max $0.7$) \\
200 & 95 & $0.80$ & $94.96$ & $0.24$ & $+0.25$ (max $0.6$) \\
500 & 95 & $0.66$ & $94.97$ & $0.19$ & $+0.09$ (max $0.3$) \\
\midrule
\multicolumn{6}{@{}l}{\textit{MSSpaceV100M}} \\
100 & 90 & $0.40$ & $90.02$ & $0.45$ & $+0.45$ (max $2.1$) \\
100 & 95 & $0.74$ & $95.02$ & $0.22$ & $+0.30$ (max $1.1$) \\
200 & 90 & $0.33$ & $90.04$ & $0.40$ & $+0.24$ (max $1.0$) \\
200 & 95 & $0.75$ & $95.01$ & $0.17$ & $+0.52$ (max $1.5$) \\
500 & 95 & $0.77$ & $95.00$ & $0.18$ & $+0.07$ (max $0.5$) \\
\bottomrule
\end{tabular}
\end{table}

\paragraph{Larger $K$.}
The conclusions hold at $K = 200$ and $500$ (\autoref{tab:predl-settings}, \autoref{tab:tvalues} and \autoref{tab:components}). The mean test recall is within $0.06$ points of $r$, and within $0.47$ points on every single split; the training and test correlations differ by at most $0.011$, and the Lagrangian is on average within $0.52\%$ of the exact optimum of~\eqref{eq:reduced}. The correlation is $0.66$--$0.80$ at $r = 95\%$, and lower at $K = 200$ and $r = 90\%$: $0.47$ on SIFT100M, $0.60$ on DEEP100M and $0.33$ on MSSpaceV100M, the settings in which beam predictor also saves the least over CBS ($1.4\%$, $2.5\%$ and $2.3\%$).

\FloatBarrier
\section{Conclusion}\label{sec:conclusion}

We studied adaptive beam search (ABS) on $\alpha$-reachable graphs, which include the graphs built by DiskANN's slow preprocessing. On $\alpha$-reachable graphs, ABS with parameter $\gamma$ returns neighbors within a factor $\tfrac{\alpha+1}{\alpha(\gamma+1)-1}$ of the true ones for any metric, and on sorted $\alpha$-reachable graphs under the $\ell_2$ metric the factor improves to $\max\bigl(\tfrac{\alpha}{\alpha(\gamma+1)-1}, \tfrac{\alpha^{2}+1}{\alpha^{2}(\gamma+1)-1}\bigr)$. These bounds unify and generalize those of Indyk and Xu~\cite{indykxu}, Gollapudi et al.~\cite{sortprune} and Al-Jazzazi et al.~\cite{adaptivebeam}, which they recover as special cases, and the general-metric bound is tight. On three 100-million-point benchmarks, with graphs built by DiskANN's fast preprocessing, which the bounds do not cover, the recall of ABS rises with $\gamma$, and the graph built with sorted pruning reaches a higher recall than the one built with shuffled pruning at every $\gamma$, in the direction that the bounds suggest.

Motivated by how ABS adapts its effort to each query, we designed beam predictor, which predicts a list size for each query from features of a short pilot search and is trained on the list sizes that minimize the total number of distance computations at a target mean recall. On the same benchmarks, with search on graphs built with DiskANN's sorted pruning rule, beam predictor lowers mean query latency on SSD-resident indices at matched recall by up to $29.6\%$ relative to classical beam search (CBS). Relative to ABS, it is faster in $11$ of $15$ settings, by up to $6.5\%$, and at most $1.4\%$ slower in the other $4$; at $95\%$ recall, its advantage over ABS on the SSD grows from $0.8$--$4.2\%$ for $K = 100$ to $4.6$--$5.6\%$ for $K = 500$. In memory, it needs $1.3$--$23.7\%$ fewer distance computations than CBS and $0.0$--$3.9\%$ fewer than ABS.

\paragraph{Limitations.} \label{subsec:limitation}
Our guarantees hold for $\alpha$-reachable and sorted $\alpha$-reachable graphs, such as those built by DiskANN's slow preprocessing, whose naive construction takes $O(n^3)$ time. The fast-preprocessing graphs used in practice and in our experiments, as well as HNSW and NSG graphs, are not known to satisfy these conditions, so the bounds do not transfer to them, and our experiments illustrate the bounds rather than test them. The bounds also concern the quality of the returned neighbors, not the running time of ABS.

Beam predictor needs an offline sweep of CBS over training queries and a target recall chosen in advance. Its gains are in mean latency; its tail latency is usually higher than that of CBS. We compare it only with CBS and ABS, not with other per-query methods (\autoref{sec:exp-setup}), and our evaluation covers three $\ell_2$ datasets of $100$ million points.

\paragraph{Future work.}
Evaluating beam predictor on more datasets, distance functions and values of $K$, and comparing it with other per-query methods, are natural next steps. A predictor that works for out-of-distribution queries is also a natural direction to work on. On the theoretical side, the second term of the approximation constant of \autoref{thm:main2} is not known to be tight, so finding a tight bound is a natural research direction.

\FloatBarrier

\section*{Acknowledgments}
We used AI assistants to help write code, analysis scripts and parts of the text, and to check the reported numbers against our experimental data. The authors reviewed and verified all content and take full responsibility for it.

\bibliographystyle{plainnat}
\bibliography{sections_arxiv/references}

\appendix
\allowdisplaybreaks
\section{Semidefinite program of Sub-case B1}\label{app:thm:mainproof}
\label{app:proofs-z0}

This appendix proves the bound of Sub-case~B1 (\autoref{subsec:z0}) for the $\ell_2$ metric on sorted $\alpha$-reachable graphs, $D(q, v)/D(q, a_i) \le \alpha/(\alpha(\gamma+1)-1)$.

Sorted $\alpha$-reachability adds the constraint $D(v, t_1) \le D(v, a_i)$, giving the optimization
\[
\max \;\frac{D(q, v)}{D(q, a_i)} \qquad \text{s.t.} \qquad
\begin{aligned}
&(\gamma+1) \cdot D(q, v) \le D(q, t_1), \\
&\alpha \cdot D(t_1, a_i) \le D(v, a_i), \\
&D(v, t_1) \le D(v, a_i).
\end{aligned}
\]
The objective and constraints depend only on ratios of $D$, so the problem is invariant under positive scaling. Without loss of generality fix $D(q, a_i) = 1$, which we further relax to $D(q, a_i) \le 1$ (any feasible solution with $D(q, a_i) < 1$ can be rescaled to increase the objective while preserving the constraints). The optimization becomes
\[
\max\; D(q, v) \qquad \text{s.t.} \qquad
\begin{aligned}
&(\gamma+1) \cdot D(q, v) \le D(q, t_1), \\
&\alpha \cdot D(t_1, a_i) \le D(v, a_i), \\
&D(v, t_1) \le D(v, a_i), \\
&D(q, a_i) \le 1.
\end{aligned}
\]

For Euclidean metric $D$ and squaring the terms of the optimization problem, we obtain a semi-definite program:
\begin{align*}
\max\;\; & \lVert q - v\rVert^2\\
\text{s.t.}\;\; & (\gamma+1)^2 \cdot \lVert q - v\rVert^2 \leq \lVert q - t_1\rVert^2,\\
& \alpha^2 \cdot \lVert a_i - t_1\rVert^2 \leq \lVert v - a_i\rVert^2,\\
& \lVert v - t_1\rVert^2 \leq \lVert v - a_i\rVert^2,\\
& \lVert q - a_i\rVert^2 \leq 1.
\end{align*}

Let the Gram matrix $X = \operatorname{Gram}(q, v, t_1, a_i) \succeq 0$. For brevity set $t = t_1$, $a = a_i$, and $\tau = \gamma + 1$.

\paragraph{SDP primal.}
\begin{align*}
p^* = \max\;\; & qq + vv - 2qv\\
\text{s.t.}\;\; & (\tau^2 - 1)qq + \tau^2 vv - tt - 2\tau^2 qv + 2qt \leq 0,\\
& (\alpha^2 - 1)aa + \alpha^2 tt - 2\alpha^2 at - vv + 2av \leq 0,\\
& tt - 2vt - aa + 2av \leq 0,\\
& qq + aa - 2aq \leq 1,\\
& X \succeq 0.
\end{align*}
\paragraph{SDP dual.}
\begin{align*}
d^* = \min\;\; & y_4\\
\text{s.t.}\;\; & \sum y_i A_i - M \succeq 0,\\
& y_i \geq 0,
\end{align*}
where
\[
X =
\begin{bmatrix}
qq & qv & qt & qa \\
vq & vv & vt & va \\
tq & tv & tt & ta \\
aq & av & at & aa
\end{bmatrix}, \qquad
M =
\begin{bmatrix}
1 & -1 & 0 & 0 \\
-1 & 1 & 0 & 0 \\
0 & 0 & 0 & 0 \\
0 & 0 & 0 & 0
\end{bmatrix},
\]
the four primal constraint matrices and their right-hand sides are
\[
A_1 = \begin{bmatrix}
\tau^{2}-1 & -\tau^{2} & 1 & 0 \\
-\tau^{2} & \tau^{2} & 0 & 0 \\
1 & 0 & -1 & 0 \\
0 & 0 & 0 & 0
\end{bmatrix}, \quad
A_2 = \begin{bmatrix}
0 & 0 & 0 & 0 \\
0 & -1 & 0 & 1 \\
0 & 0 & \alpha^{2} & -\alpha^{2} \\
0 & 1 & -\alpha^{2} & \alpha^{2}-1
\end{bmatrix},
\]
\[
A_3 = \begin{bmatrix}
0 & 0 & 0 & 0 \\
0 & 0 & -1 & 1 \\
0 & -1 & 1 & 0 \\
0 & 1 & 0 & -1
\end{bmatrix}, \quad
A_4 = \begin{bmatrix}
1 & 0 & 0 & -1 \\
0 & 0 & 0 & 0 \\
0 & 0 & 0 & 0 \\
-1 & 0 & 0 & 1
\end{bmatrix},
\]
\[
b_1 = b_2 = b_3 = 0,\; b_4 = 1.
\]

Substituting the $A_i$'s, the dual feasibility constraint is equivalent to
\[
\fitwidth{$\displaystyle \setlength{\arraycolsep}{3pt}
Y =
\begin{bmatrix}
(\tau^{2}-1) y_1 + y_4 - 1 & -\tau^{2} y_1 + 1 & y_1 & -y_4 \\
-\tau^{2} y_1 + 1 & \tau^{2} y_1 - y_2 - 1 & -y_3 & y_2 + y_3 \\
y_1 & -y_3 & -y_1 + \alpha^{2} y_2 + y_3 & -\alpha^{2} y_2 \\
-y_4 & y_2 + y_3 & -\alpha^{2} y_2 & (\alpha^{2}-1) y_2 - y_3 + y_4
\end{bmatrix} \;\succeq\; 0.$}
\]

\paragraph{Dual certificate.}
We exhibit a dual-feasible point:
\begin{gather*}
y_1 = \frac{\alpha}{\tau(\alpha\tau - 1)}, \quad
y_2 = \frac{\alpha}{(\alpha\tau - 1)\bigl(\alpha(\tau^{2}+1) - \tau\bigr)},\\
y_3 = \frac{\alpha}{\alpha(\tau^{2}+1) - \tau}, \quad
y_4 = \left(\frac{\alpha}{\alpha\tau - 1}\right)^{2}.
\end{gather*}
All four values are positive in the parameter range $\tau \geq 1$ that is relevant here. Substituting gives
\[
\fitwidth{$\displaystyle \setlength{\arraycolsep}{3pt}
Y =
\left[
\begin{array}{cccc}
\displaystyle \frac{\alpha\tau^{2}+\alpha-\tau}{\tau(\alpha\tau-1)^{2}} &
\displaystyle -\frac{1}{\alpha\tau-1} &
\displaystyle \frac{\alpha}{\tau(\alpha\tau-1)} &
\displaystyle -\frac{\alpha^{2}}{(\alpha\tau-1)^{2}} \\[10pt]
\displaystyle -\frac{1}{\alpha\tau-1} &
\displaystyle \frac{\tau}{\alpha\tau^{2}+\alpha-\tau} &
\displaystyle -\frac{\alpha}{\alpha(\tau^{2}+1)-\tau} &
\displaystyle \frac{\alpha^{2}\tau}{(\alpha\tau-1)(\alpha\tau^{2}+\alpha-\tau)} \\[10pt]
\displaystyle \frac{\alpha}{\tau(\alpha\tau-1)} &
\displaystyle -\frac{\alpha}{\alpha(\tau^{2}+1)-\tau} &
\displaystyle \frac{\alpha^{2}}{\tau(\alpha\tau^{2}+\alpha-\tau)} &
\displaystyle -\frac{\alpha^{3}}{(\alpha\tau-1)\bigl(\alpha(\tau^{2}+1)-\tau\bigr)} \\[10pt]
\displaystyle -\frac{\alpha^{2}}{(\alpha\tau-1)^{2}} &
\displaystyle \frac{\alpha^{2}\tau}{(\alpha\tau-1)(\alpha\tau^{2}+\alpha-\tau)} &
\displaystyle -\frac{\alpha^{3}}{(\alpha\tau-1)\bigl(\alpha(\tau^{2}+1)-\tau\bigr)} &
\displaystyle \frac{\alpha^{4}\tau}{(\alpha\tau-1)^{2}(\alpha\tau^{2}+\alpha-\tau)}
\end{array}
\right].$}
\]
Setting
\[
U =
\begin{bmatrix}
\displaystyle \frac{1}{\alpha\tau-1}\sqrt{\dfrac{\alpha(\tau^{2}+1)-\tau}{\tau}} \\[10pt]
\displaystyle -\sqrt{\dfrac{\tau}{\alpha(\tau^{2}+1)-\tau}} \\[10pt]
\displaystyle \frac{\alpha}{\tau}\sqrt{\dfrac{\tau}{\alpha(\tau^{2}+1)-\tau}} \\[10pt]
\displaystyle -\frac{\alpha^{2}}{\alpha\tau-1}\sqrt{\dfrac{\tau}{\alpha(\tau^{2}+1)-\tau}}
\end{bmatrix}
\]
one verifies by direct computation that $Y = U U^{\top}$; hence $Y \succeq 0$ and the chosen $(y_1, y_2, y_3, y_4)$ is dual-feasible. By weak duality,
\[
p^{\star} \;\le\; d^{\star} \;\le\; y_4 \;=\; \left(\frac{\alpha}{\alpha(\gamma+1)-1}\right)^{2},
\]
and taking square roots recovers the bound $D(q, v) / D(q, a_i) \le \alpha / (\alpha(\gamma+1) - 1)$ claimed in Sub-case~B1 (\autoref{subsec:z0}).

\section{Tail latency and per-query recall}\label{app:tails}

\autoref{tab:tails} complements \autoref{tab:ssd}. For the same runs and the same matched settings of CBS and ABS, it gives the $99.9$th-percentile latency of each method, averaged over the five runs, and the median and standard deviation of per-query recall@$K$ over the test queries, which are the same in every run.

\begin{table}[!htbp]
\centering
\small
\caption{SSD index, sorted graph, same runs and matched settings as \autoref{tab:ssd}: $99.9$th-percentile latency (ms, mean over the five runs), and median and standard deviation (Std) of per-query recall@$K$ (\%) over the test queries. BP: beam predictor.}
\label{tab:tails}
\setlength{\tabcolsep}{4pt}
\begin{tabular}{@{}rrrrrrrrrrr@{}}
\toprule
 & & \multicolumn{3}{c}{$99.9$th-pct.\ latency} & \multicolumn{3}{c}{Median recall} & \multicolumn{3}{c}{Std of recall} \\
\cmidrule(lr){3-5}\cmidrule(lr){6-8}\cmidrule(l){9-11}
$K$ & $r$ & CBS & ABS & BP & CBS & ABS & BP & CBS & ABS & BP \\
\midrule
\multicolumn{11}{@{}l}{\textit{SIFT100M}} \\
100 & 90 & $12.6$ & $13.3$ & $16.8$ & $94.0$ & $93.0$ & $92.0$ & $10.8$ & $10.2$ & $9.1$ \\
100 & 95 & $17.4$ & $29.4$ & $36.7$ & $98.0$ & $97.0$ & $97.0$ & $6.9$ & $6.1$ & $5.2$ \\
200 & 90 & $20.6$ & $19.8$ & $20.4$ & $93.0$ & $93.0$ & $93.0$ & $10.0$ & $9.9$ & $9.6$ \\
200 & 95 & $29.3$ & $33.8$ & $46.4$ & $97.5$ & $97.0$ & $96.5$ & $6.4$ & $5.8$ & $4.7$ \\
500 & 95 & $41.2$ & $44.2$ & $65.8$ & $97.2$ & $97.0$ & $96.4$ & $5.9$ & $5.7$ & $4.8$ \\
\midrule
\multicolumn{11}{@{}l}{\textit{DEEP100M}} \\
100 & 90 & $13.4$ & $17.0$ & $17.7$ & $94.0$ & $94.0$ & $93.0$ & $11.8$ & $11.1$ & $10.4$ \\
100 & 95 & $24.7$ & $35.6$ & $40.5$ & $98.0$ & $98.0$ & $97.0$ & $7.7$ & $6.8$ & $6.3$ \\
200 & 90 & $23.5$ & $25.5$ & $22.9$ & $93.5$ & $93.5$ & $93.0$ & $11.0$ & $10.7$ & $10.1$ \\
200 & 95 & $35.5$ & $41.9$ & $58.9$ & $98.0$ & $97.5$ & $97.0$ & $7.2$ & $6.5$ & $5.9$ \\
500 & 95 & $82.0$ & $58.9$ & $94.5$ & $97.6$ & $97.2$ & $96.6$ & $6.7$ & $6.3$ & $5.3$ \\
\midrule
\multicolumn{11}{@{}l}{\textit{MSSpaceV100M}} \\
100 & 90 & $15.1$ & $17.6$ & $16.3$ & $96.0$ & $96.0$ & $96.0$ & $15.8$ & $15.1$ & $15.3$ \\
100 & 95 & $33.2$ & $122.2$ & $52.5$ & $99.0$ & $98.0$ & $98.0$ & $9.5$ & $8.4$ & $8.1$ \\
200 & 90 & $20.0$ & $21.5$ & $20.2$ & $95.5$ & $95.0$ & $95.0$ & $13.8$ & $13.5$ & $13.5$ \\
200 & 95 & $44.7$ & $136.2$ & $82.9$ & $98.5$ & $98.0$ & $97.5$ & $8.5$ & $7.6$ & $7.2$ \\
500 & 95 & $95.5$ & $265.8$ & $163.4$ & $98.2$ & $97.6$ & $97.4$ & $7.8$ & $6.9$ & $6.5$ \\
\bottomrule
\end{tabular}
\end{table}

\end{document}